\documentclass{article}

\PassOptionsToPackage{numbers, compress}{natbib}

\usepackage[final]{neurips_data_2022}

\usepackage{amsmath}
\usepackage{graphicx}
\usepackage[utf8]{inputenc} 
\usepackage[T1]{fontenc}    
\usepackage{hyperref}       
\usepackage{url}            
\usepackage{booktabs}       
\usepackage{amsfonts}       
\usepackage{nicefrac}       
\usepackage{subfigure}
\usepackage{tabularx}
\usepackage{microtype}      
\usepackage{xcolor}         
\usepackage[toc,page]{appendix}
\usepackage{amsmath}
\usepackage{wrapfig,lipsum,booktabs}
\usepackage[para,online,flushleft]{threeparttable}
\usepackage{enumitem}
\usepackage{multirow}
\usepackage{wrapfig}
\usepackage{colortbl,xcolor}

\let\llncssubparagraph\subparagraph
\let\subparagraph\paragraph
\usepackage[compact]{titlesec}
\let\subparagraph\llncssubparagraph
\titlespacing{\section}{0pt}{3ex plus .2ex minus 1ex}{2ex minus 1ex}
\titlespacing{\subsection}{0pt}{2ex plus .2ex minus 1ex}{1ex minus 1ex}

\setlist[itemize]{align=parleft,left=0pt}

\definecolor{azure(colorwheel)}{rgb}{0.0, 0.5, 1.0}
\definecolor{nicegreen}{rgb}{0.0, 0.7, 0.1}
\definecolor{clova}{rgb}{0.24, 0.63, 0.33}
\definecolor{customgray}{rgb}{0.9, 0.9, 0.9}
\definecolor{pink}{cmyk}{0, 0.7808, 0.4429, 0.1412}
\definecolor{amethyst}{rgb}{0.6, 0.4, 0.8}
\definecolor{black}{rgb}{0.0, 0.0, 0.0}
\definecolor{white}{rgb}{1.0, 1.0, 1.0}
\definecolor{red}{rgb}{0.9, 0.0, 0.}

\newcolumntype{g}{>{\columncolor{customgray}}c}
\newcolumntype{z}{>{\columncolor{customgray}}l}
\newcolumntype{?}[1]{!{\vrule width #1}}

\renewcommand{\paragraph}[1]{\vspace{1mm}\noindent\textbf{#1.}\,\,}

\usepackage{xspace}

\newcommand{\be}{\begin{eqnarray}}
\newcommand{\ee}{\end{eqnarray}}
\newcommand{\bee}{\begin{eqnarray*}}
\newcommand{\eee}{\end{eqnarray*}}

\newcommand{\matrixb}{\left[ \begin{array}}
\newcommand{\matrixe}{\end{array} \right]}   

\title{ComMU: Dataset for Combinatorial Music Generation}

\author{%
    Lee Hyun\thanks{Equal Contribution} \ \thanks{Work done at Pozalabs, now at POSTECH EE} \\
    Pozalabs \\
    \texttt{hyun@pozalabs.com} \\
    \And
    Taehyun Kim\footnotemark[1] \\
    Pozalabs, Yonsei Univ. \\
   \texttt{taehyun@pozalabs.com} \\
   \texttt{kimth0101@yonsei.ac.kr} \\
   \AND
   Hyolim Kang \\
   Yonsei Univ. \\
   \texttt{hyolimkang@yonsei.ac.kr} \\
   \And
   Minjoo Ki \\
   Yonsei Univ. \\
   \texttt{minjoo@yonsei.ac.kr} \\
   \And
   Hyeonchan Hwang \\
   Pozalabs \\
   \texttt{hyeonchan@pozalabs.com} \\
   \AND
   Kwanho Park \\
   Pozalabs \\
   \texttt{kwanho@pozalabs.com} \\
   \And
   Sharang Han \\
   Pozalabs \\
   \texttt{sharang@pozalabs.com} \\
   \And
   Seon Joo Kim \\
   Pozalabs, Yonsei Univ. \\
   \texttt{seonjoo@pozalabs.com} \\
   \texttt{seonjookim@yonsei.ac.kr} \\
}

\begin{document}

\maketitle

\begin{abstract}
    Commercial adoption of automatic music composition requires the capability of generating diverse and high-quality music suitable for the desired context (e.g., music for romantic movies, action games, restaurants, etc.). In this paper, we introduce \emph{combinatorial music generation}, a new task to create varying background music based on given conditions. \emph{Combinatorial music generation} creates short samples of music with rich musical metadata, and combines them to produce a complete music. In addition, we introduce \emph{ComMU}, the first symbolic music dataset consisting of short music samples and their corresponding 12 musical metadata for \emph{combinatorial music generation}. Notable properties of ComMU are that (1) dataset is manually constructed by professional composers with an objective guideline that induces regularity, and (2) it has 12 musical metadata that embraces composers' intentions. Our results show that we can generate diverse high-quality music only with metadata, and that our unique metadata such as track-role and extended chord quality improves the capacity of the automatic composition. We highly recommend watching our video before reading the paper (\url{https://pozalabs.github.io/ComMU/}).


\end{abstract}

\section{Introduction} \label{intro}
    
Although musical composition is a creative process, algorithmic approaches for automatic music composition have been continuously studied~\cite{nierhaus2009historical, giraudo2021generation}.
Recently, deep learning has shown great potential in composition.
\citet{bretan2016} and \citet{45871} have introduced deep sequence models into generating music sequences.
After the seminal works, prior works \citep{huang2018music, payne2019musenet, jiang2020transformer, muhamed2021symbolic} have proposed conditional music generation with initial sequences and musical metadata using language models.
Such deep models have improved the quality of the composition and can create authentic music.

\begin{figure*}[!t]
\label{fig:intro}
\centering
\subfigure[$\mathtt{stage 1}$]{
\includegraphics[width=0.38\textwidth, height=31mm]{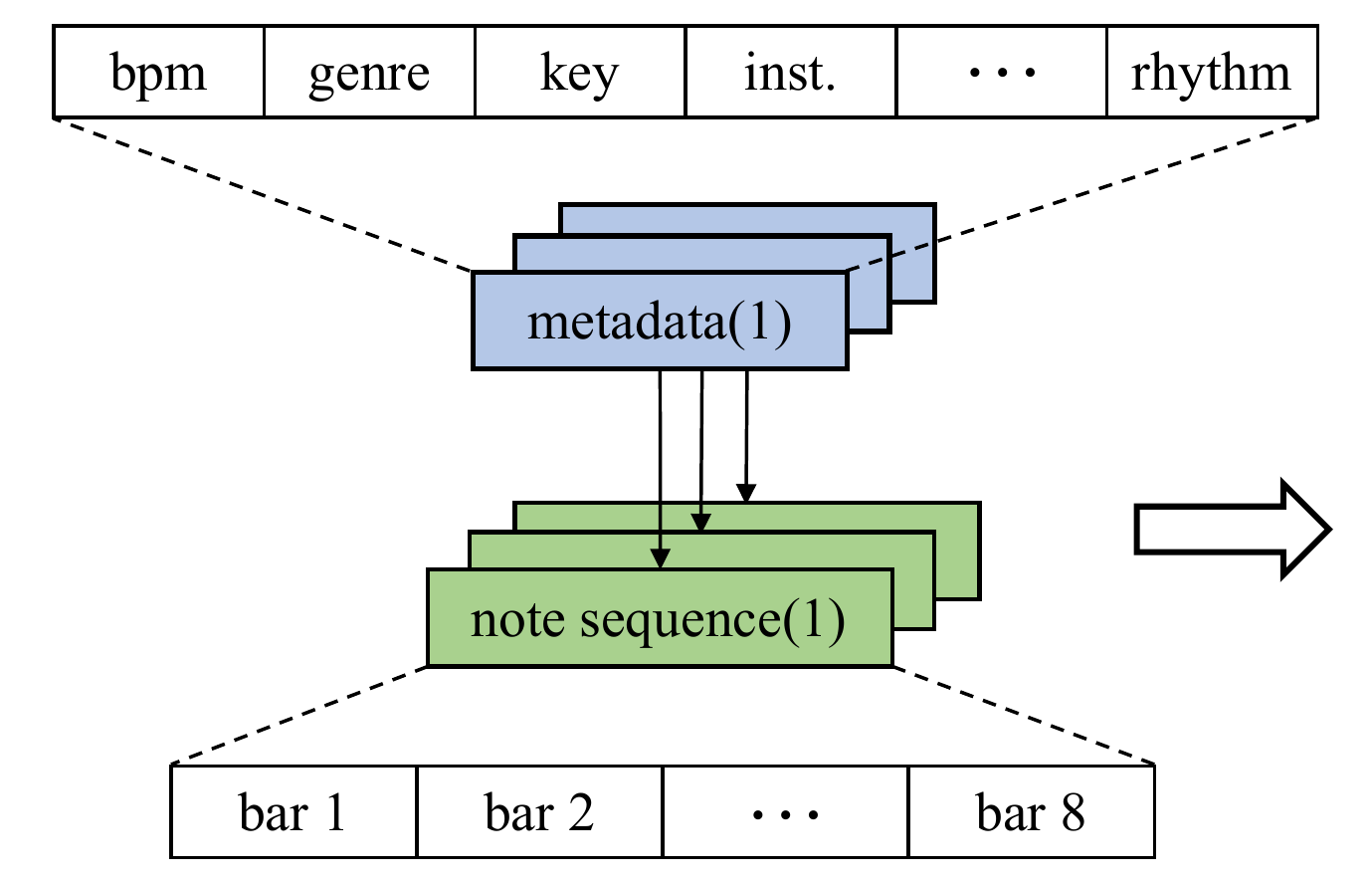}
\label{fig:intro_a}
}
\subfigure[$\mathtt{stage 2}$]{
\includegraphics[width=0.58\textwidth, height=31mm]{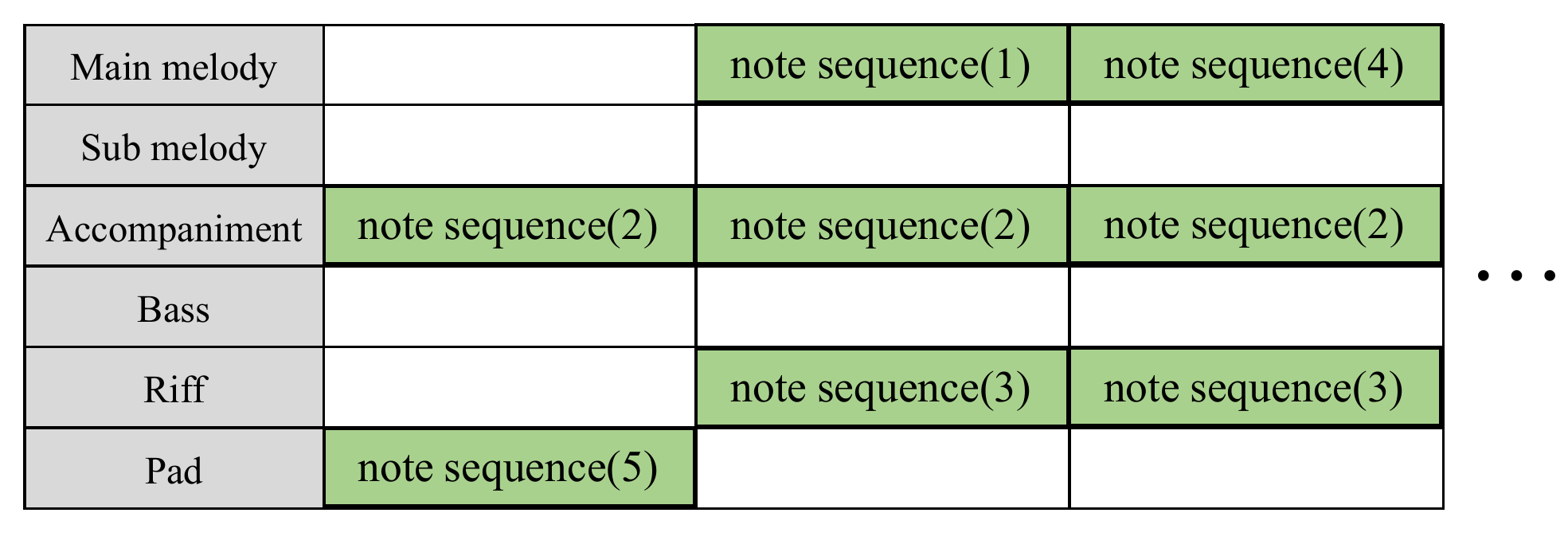}
\label{fig:intro_b}
}
\caption{\textbf{The whole process of \emph{Combinatorial Music Generation}}. In $\mathtt{stage 1}$, a note sequence ( green box) is generated from a set (blue box) of metadata. $\mathtt{Stage 2}$ then combines note sequences generated from several metadata sets to create one completed music. The number of bars for a note sequence and the number of note sequences for a complete piece of music can be flexible (\emph{ComMU} mostly has a note sequence of 4, 8, and 16 bars). 
In this work, we focus on solving $\mathtt{stage 1}$.}
\vspace{-3mm}
\end{figure*}
While the models continue to improve and generate authentic music, creating music on a commercially usable level still remains as an issue. 
Popular music is mostly a \emph{homophony} in the sense that one track is in charge of the main melody while remaining tracks harmonically support the main track~\cite{sadie2001dictionary}.
As the main melody and accompanying tracks are strictly separated, the common practice of composing the homophony music is \emph{combinatorial}.
In other words, tracks including the main melody and its accompaniments are separately generated and combined afterwards.
Despite this lazy combination, the dissonance between them is prevented by the chord-conditioned generation of each track.
In this context, we propose \emph{combinatorial music generation}, which mimics the human composition convention on homophony music.
Figure~\ref{intro} shows the overall process of generating combinatorial music -- track-level note sequences are created with a set of metadata, which are combined later to create a complete piece of music.


To generate commercially usable music, it is essential to have a detailed control over the generating process. For example, when a movie director requests fast and tense music suitable for an action scene, composers tend to pre-set proper musical metadata that affects the mood of the music — chord progression, key, genre, and rhythm — then compose music based on the metadata.
From this point of view, we design $\mathtt{stage 1}$ (Figure~\ref{fig:intro_a}) that generates note sequences, the elements for combinatorial music generation, with rich musical metadata to embrace composer's intention and harmony.

All in all, combinatorial music generation must be harmonized by vertically stacking note sequences within the homophony scheme and capture the intention of composition. Therefore, generating note sequences under elaborate musical controls such as track-role, instrument, and chord progression is necessary.
However, previous symbolic music datasets~\cite{raffel2016learning, donahue2019lakhnes, hawthorne2018enabling, dorfer2018learning, ferreira2020computer, kong2020giantmidi} do not have enough metadata for sophisticated control.
To tackle this issue, we present \emph{ComMU}, a symbolic music dataset containing 11,144 MIDI samples that consist of short note sequences that are manually composed by professional composers with its corresponding 12 metadata (bpm, genre, key, instrument, track-role, time signature, pitch range, number of measures, chord progression, min/max velocity, and rhythm). 

Our rich metadata can embrace desired musical conditions required for combinatorial music generation. It nicely controls the generated music by the given metadata and creates diverse music close to human creativity, leveraging numerous combinations of metadata at $\mathtt{stage 1}$ (Figure~\ref{fig:intro_a}) and note sequences at $\mathtt{stage 2}$ (Figure~\ref{fig:intro_b}).
While previous tasks on music generation have centered on music continuation~\cite{shih2022theme}, reconstruction~\cite{von2022figaro}, style transfer~\cite{wu2021musemorphose}, or creating music with a few metadata~\cite{ens2020mmm, dong2018musegan, zhu2018xiaoice, zhu2020pop, hung2021emopia, di2021video}, we focus on composing diverse music with abundant metadata, similar to the way a composer generates music.

In this paper, we focus on $\mathtt{stage 1}$, the combinatorial music generation with the ComMU dataset.
We evaluate the controllability, fidelity, and the diversity of generated note sequences. Our results show that (1) combination of multiple metadata can generate diverse and high-quality music with an auto-regressive language model, (2) the unique metadata (e.g., extended chord quality, track-role) improves the capacity and flexibility of the automatic composition.

Overall, the main contributions of this paper are:
\begin{itemize}
\item We propose the combinatorial music generation task with the ComMU dataset for the industry-level automatic music composition. Diverse and high-quality music is created with our framework and dataset. 
\item ComMU is the first symbolic music dataset manually created by professional composers for automatic music composition with 12 metadata.
\item We show that our unique metadata such as track-role and extended chord quality are essential musical metadata for human-like composition, as they play a crucial role in expressing the comprehensive intention of the composer.
\end{itemize} 

\section{Related work}
    \textbf{Conditional music generation.} $\mathtt{Stage 1}$ of the combinatorial music generation task can be considered as an extension of conditional music generation in that it generates a track with a given set of metadata.
Many preceding works share similar scheme; for instance, MuseNet~\cite{payne2019musenet} creates music sequences in compliance with the given music style or instruments, and FIGARO~\cite{von2022figaro} generates music with a few metadata that are extracted from a reference music.
There are models~\cite{Yang2017MidiNetAC, tan2019chordal, Hakimi2020BebopNetDN} that can generate note sequences based on a given chord progression.
They can produce music that fits the chord, but do not convey other important metadata such as rhythm and instrument.

MMM~\cite{ens2020mmm} is the closest task to ours, which takes instruments, bpm, and the number of bars as conditions and produces multiple instrument-tracks.
However, MMM differs from our task in that the generated track cannot be combined into the homophony scheme with coherence, because it cannot take the track-role and the chord progression as conditions.
It is rather specialized towards resampling or inpainting with a given music.

\begin{table}
  \caption{Comparison of ComMU to recent MIDI datasets with various metadata. We compare ComMU to other MIDI dataset on 4 types of metadata: genre, instrument, track-role, and chord progression.}
  \vspace{3mm}
  \centering
  
  \label{sample-table}
  \resizebox{0.95\columnwidth}{!}{
  \begin{tabular}{l|cccc}
    \toprule
    Dataset & Genre & Instrument & Track-role & Chord progression \\
    \midrule
    Lakh MIDI~\cite{raffel2016learning} & \checkmark & \checkmark & - & - \\
    MAESTRO~\cite{hawthorne2018enabling} & \checkmark & (\checkmark)$^{1}$ & - & - \\
    MSMD~\cite{dorfer2018learning} & - & (\checkmark)$^{1}$ & - & \checkmark \\
    ADL Piano MIDI~\cite{ferreira2020computer} & \checkmark & (\checkmark)$^{1}$ & - & - \\
    GiantMIDI-Piano~\cite{kong2020giantmidi} & - & (\checkmark)$^{1}$ & - & \checkmark \\
    EMOPIA~\cite{hung2021emopia} & - & (\checkmark)$^{1}$ & - & \checkmark \\
    \hline
    \textbf{ComMU}(Ours) & \checkmark & \checkmark & \checkmark & \checkmark \\
    \bottomrule
  \end{tabular}}
  \begin{tablenotes}
    \item [1] include only one instrument.
    \item
  \end{tablenotes}
\vspace{-1mm}
\end{table}

\textbf{Symbolic music dataset.} Various music datasets with metadata have been introduced with the development of conditional music generation. Although there are various forms of expressing music datasets, such as MIDI, audio, and piano-roll, we focus on MIDI-based datasets with metadata for the comparison with ComMU.  Lakh MIDI Dataset~\cite{raffel2016learning}, MAESTRO~\cite{hawthorne2018enabling}, and ADL Piano MIDI~\cite{ferreira2020computer} have various meta-information such as genre, instrument, key, and time signature, but there is no chord information for the harmony of track-level composition.
On the other hand, MSMD~\cite{dorfer2018learning}, GiantMIDI-Piano~\cite{kong2020giantmidi}, and EMOPIA~\cite{hung2021emopia} have chord information that can infuse harmony, but lack metadata such as genre, and rhythm that can reflect the intention of composition.

Aside from the fact that existing symbolic music datasets do not have rich metadata, there are no datasets with track-role information. Although some datasets~\cite{raffel2016learning, SiSEC16, rafii2017musdb18} define track as instruments, ComMU is the first dataset to separate the instrument from the track-role, allowing combinatorial music generation. We list major differences between ComMU with other datasets in Table~\ref{sample-table}.

\section{ComMU dataset}
    \subsection{Dataset collection} \label{subsection:dataset_collection}

\begin{wraptable}{r}{0.45\linewidth}
\vspace{-1mm}
\caption{Basic information of ComMU.}
\begin{threeparttable}
\vskip 0.015in
\begin{center}
\begin{small}
\begin{sc}
\begin{tabular}{l|r}
\toprule
\# samples & 11,144\\
\# notes & 526,612\\
\midrule
Types of Audio keys & 24\\
Types of Instruments & 37\\
Types of Genres & 2\\
Types of Rhythm & 2\\
Types of Pitch Range & 7\\
Types of Track-Role & 6\\
Types of Time Signature & 3\\
Types of Num-Measures & 3\\
\midrule
Range of BPM & 35-160$^{1}$\\
Range of Min Velocity & 2-127$^{2}$\\
Range of Max Velocity & 2-127$^{2}$\\
\bottomrule
\end{tabular}
\end{sc}
\begin{tablenotes}
  \item[1] The range of representation is 5-200.
  \item[2] Velocity 1 is for keyswitch note which is a key pressing when changing the playing type for each instrument. 
  \end{tablenotes}
\end{small}
\end{center}
  \end{threeparttable}
\vskip -0.4in
\label{table:basic-info}
\end{wraptable}

ComMU has 11,144 MIDI samples that consist of short note sequences with their corresponding 12 metadata. Basic information of ComMU dataset is given in Table~\ref{table:basic-info}. 

\begin{figure}
\centering
\includegraphics[width=1.0\textwidth]{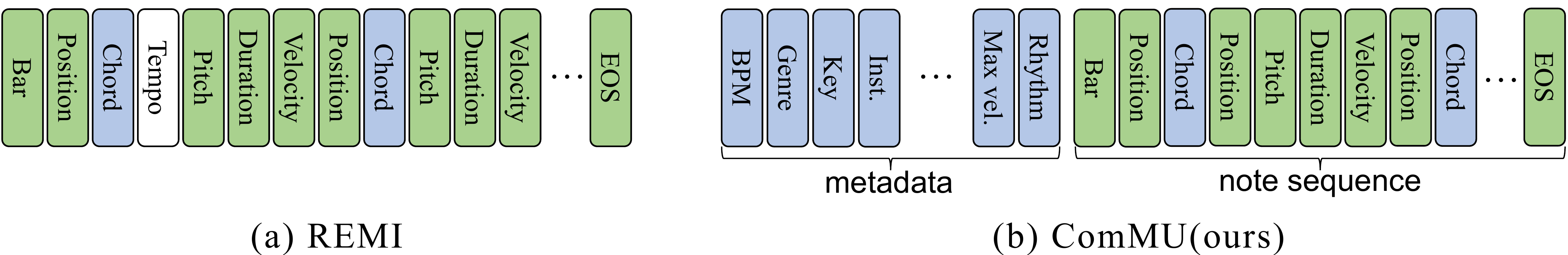}
\caption{Illustration of REMI and ComMU representation. ComMU differs from REMI in that it has metadata, extended chords and increased note resolution, and eliminates tempo token.}
\label{data_represen}
\vspace{-1mm}
\end{figure}

Fourteen professional composers have manually made MIDI samples with metadata for 6 months. In detail, the composers were split into two teams. The first team created composition guidelines for certain genres and moods by finding and analyzing reference music. Each guideline then becomes an instruction (i.e., metadata) for composing samples, such as instruments, track-role, chord progressions, and pitch range available for each sample. The second team composed MIDI samples in accordance with each guideline, taking a total of 6 months to produce 11,144 music samples. The music made with these objective guidelines acquires regularity, suitable for the machine learning system.

\subsection{Pre-processing and representation} \label{dataset-representation}
When creating the dataset, deciding on the appropriate data representation is one of the most important considerations. Our representation is based on REMI~\cite{huang2020pop}, which encodes music samples in a token-based fashion. Throughout 12 metadata, REMI supports representing chord progression among 60 patterns, but we extend them to 108 patterns to hold diverse chord quality. Other 11 metadata are placed before the REMI representation. We also remove the tempo token as the tempo change never happens in our samples due to their short length. In addition, we increase the resolution of the position and the duration token from 32 notes in REMI to 128 notes to improve the quality of the result. See Appendix~\ref{appendix:resolution} for an empirical study of the representation resolution. We compare REMI with our revised representation in Figure~\ref{data_represen}.

Unlike other metadata, the chord progression cannot be encoded as a single token due to its sequential form. Therefore, we encode the chord progression into a note sequence with position tokens. See Appendix~\ref{appendix: preprocessing} for more details about pre-processing and data representation.
\subsection{Metadata}
We provide the definition of 3 metadata which can be relatively ambiguous. Others including bpm, key, instrument, time signature, pitch range, number of measure, min/max velocity, and rhythm are explained in Appendix~\ref{appendix:metadata}.

\textbf{Genre.} We chose the new age and the cinematic genres for our dataset, which are often used in background music. We define the new age as a melodious genre mainly consisting of keyboard instruments and small-scale instruments such as acoustic instruments. The cinematic genre is a large-scale genre with orchestra, especially involving classical instruments such as string ensembles in charge of the melody and the accompaniment.

\textbf{Track role.} It is a classification of what role the created note sequences have in a multi-track music. We divide multi-track into main melody, sub melody, accompaniment, bass, pad, and riff.





\textbf{Chord progression.} Chord progression is the set of chords that are used in a sample. Extended chord quality leads to the improvement in harmonic performance and diversity by letting the model learn various possible melodies associated with the same chord progression.




\subsection{Data analysis} 
\label{data_analysis}
In order to create music that is well-controlled only by the metadata, it is beneficial to have a distinguishable distribution of note sequences according to the metadata in ComMU. We extract MIDI-based features and analyze the distribution according to the metadata to observe the correlation between the metadata and the note sequences. In addition, we analyze the correlation as a heatmap, showing that a transformer~\cite{vaswani2017attention} that covers the long-term dependency is suitable as a baseline. Among various options in our analysis, we present relevant features below.

\begin{figure*}[!t]
\label{fig:intro}
\centering
\subfigure[Note density - Track-role]{
\includegraphics[width=0.48\textwidth, height=32mm]{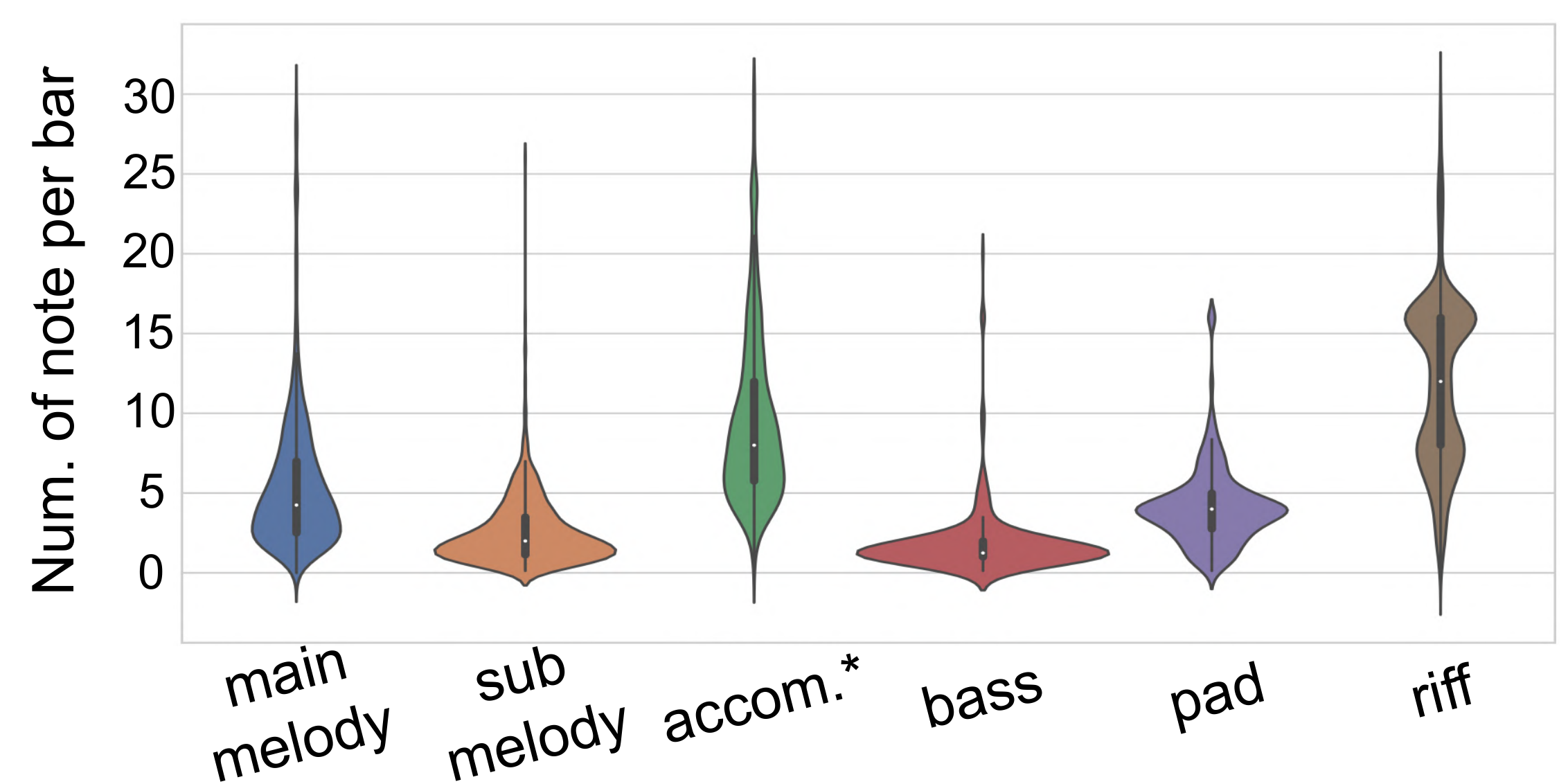}
}
\subfigure[Note length - Track-role]{
\includegraphics[width=0.48\textwidth, height=32mm]{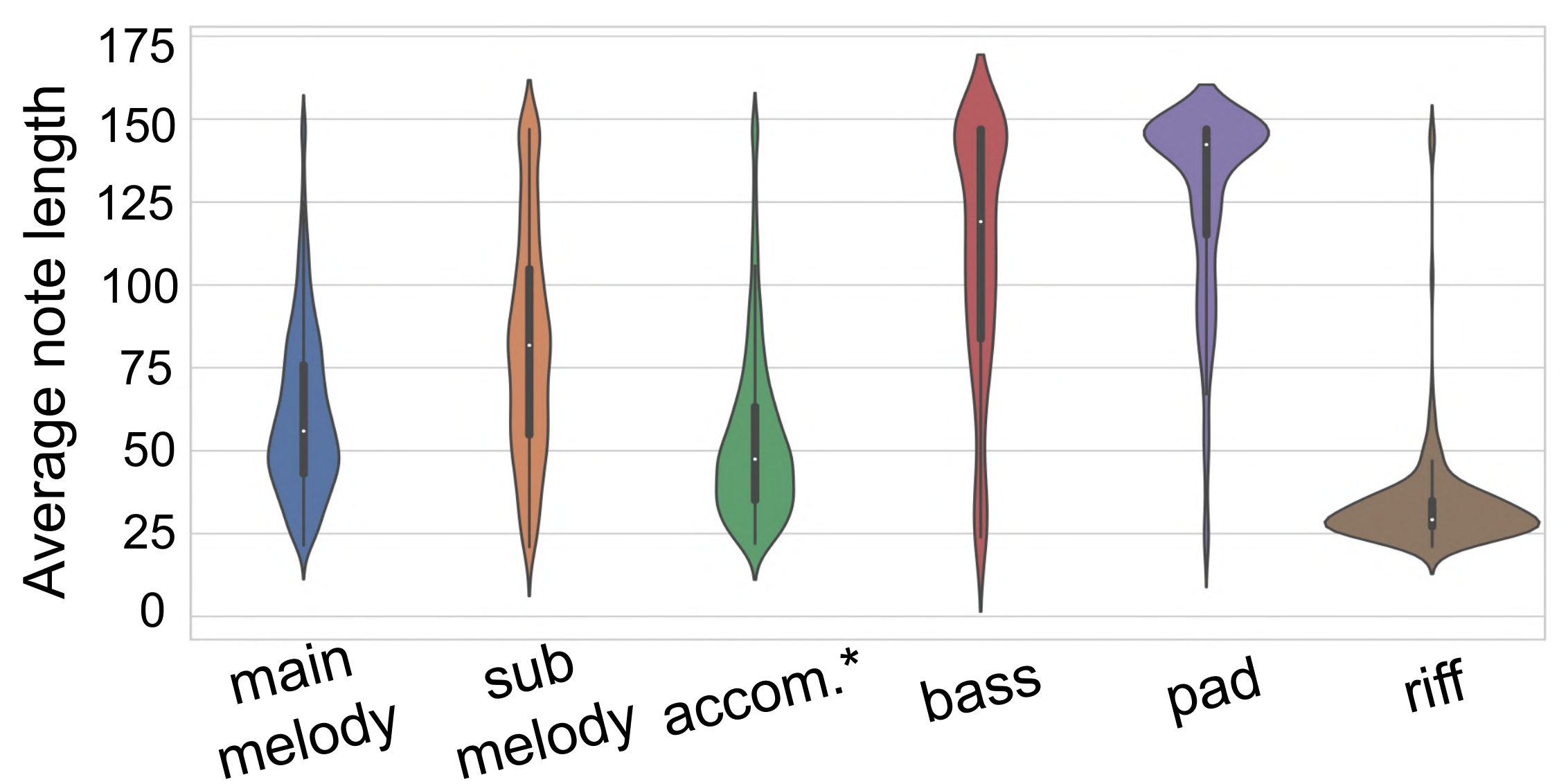}
}
\caption{Illustration of the distribution in note density and note length, according to the track-role. Depending on the track-role, the shape of the corresponding notes varies. \emph{*accom.: accompaniment.}}
\label{track_notes}
\vspace{-3mm}
\end{figure*}

\textbf{Analysis between metadata and notes.} 
We measure the characteristics of the note sequence by note density and length~\cite{hung2021emopia}. Note density is the average of the number of notes appearing in one bar, and note length is the average of the lengths of all notes in a sample.
Figure~\ref{track_notes} shows the distribution of note density and length according to the track category. Melody and accompaniment have short note lengths, whereas bass and pad have relatively long notes. Note density also shows a well-distinguished distribution of melody/accompaniment, bass/pad, and riff groups. This means that the musical dynamics of the melody/accompaniment are relatively strong, while the bass/pad have weak and stable notes. The riff track has a high density as the accompaniment track or higher. Considering it with the length, we can infer that the sound pattern with repeated short notes is the characteristic of the riff track.


\begin{figure*}[!t]
\label{fig:intro}
\centering
\subfigure[Instrument - Track-role]{
\includegraphics[width=0.48\textwidth, height=38mm]{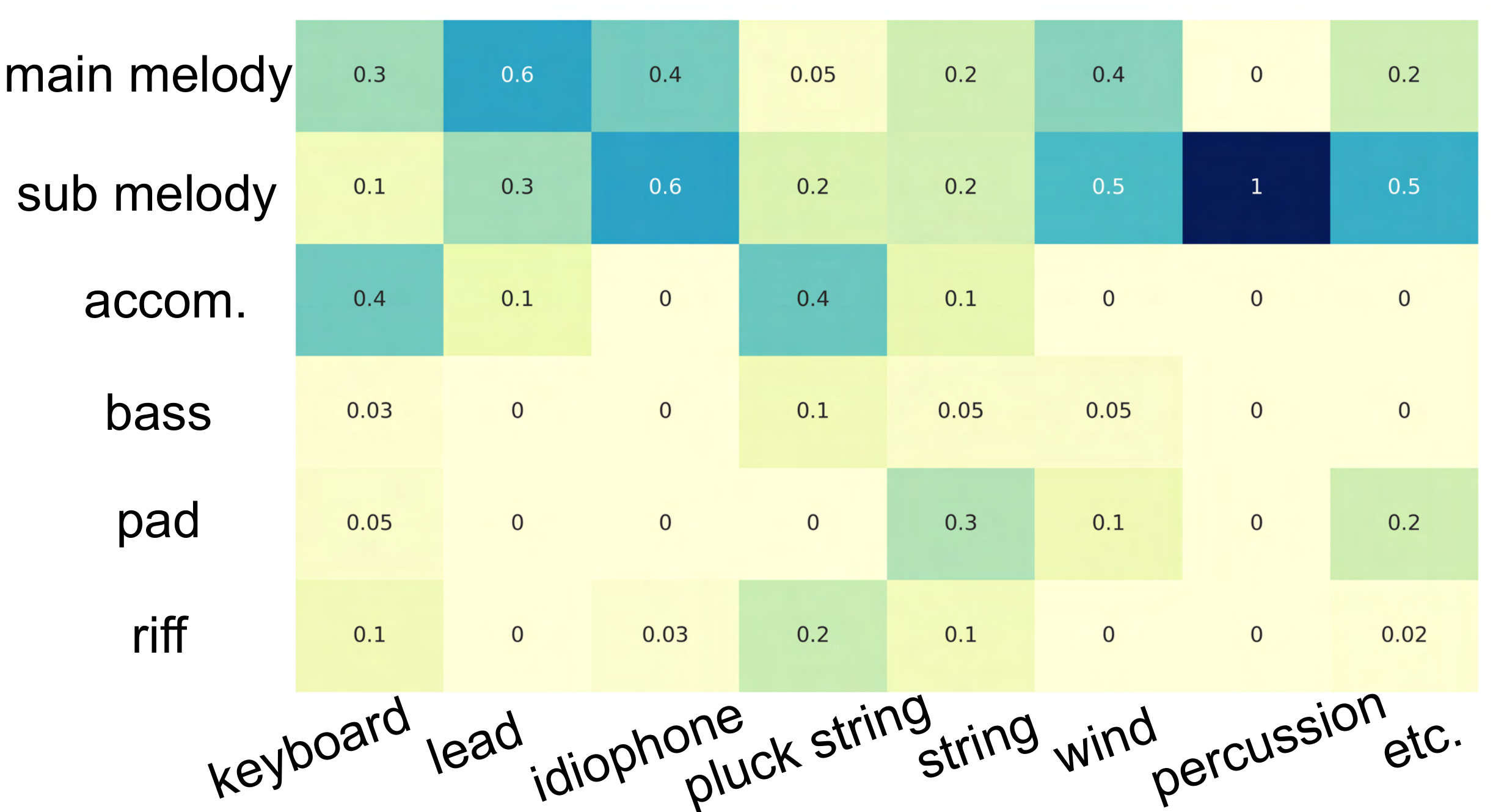}
}
\subfigure[Pitch range - Track-role]{
\includegraphics[width=0.48\textwidth, height=38mm]{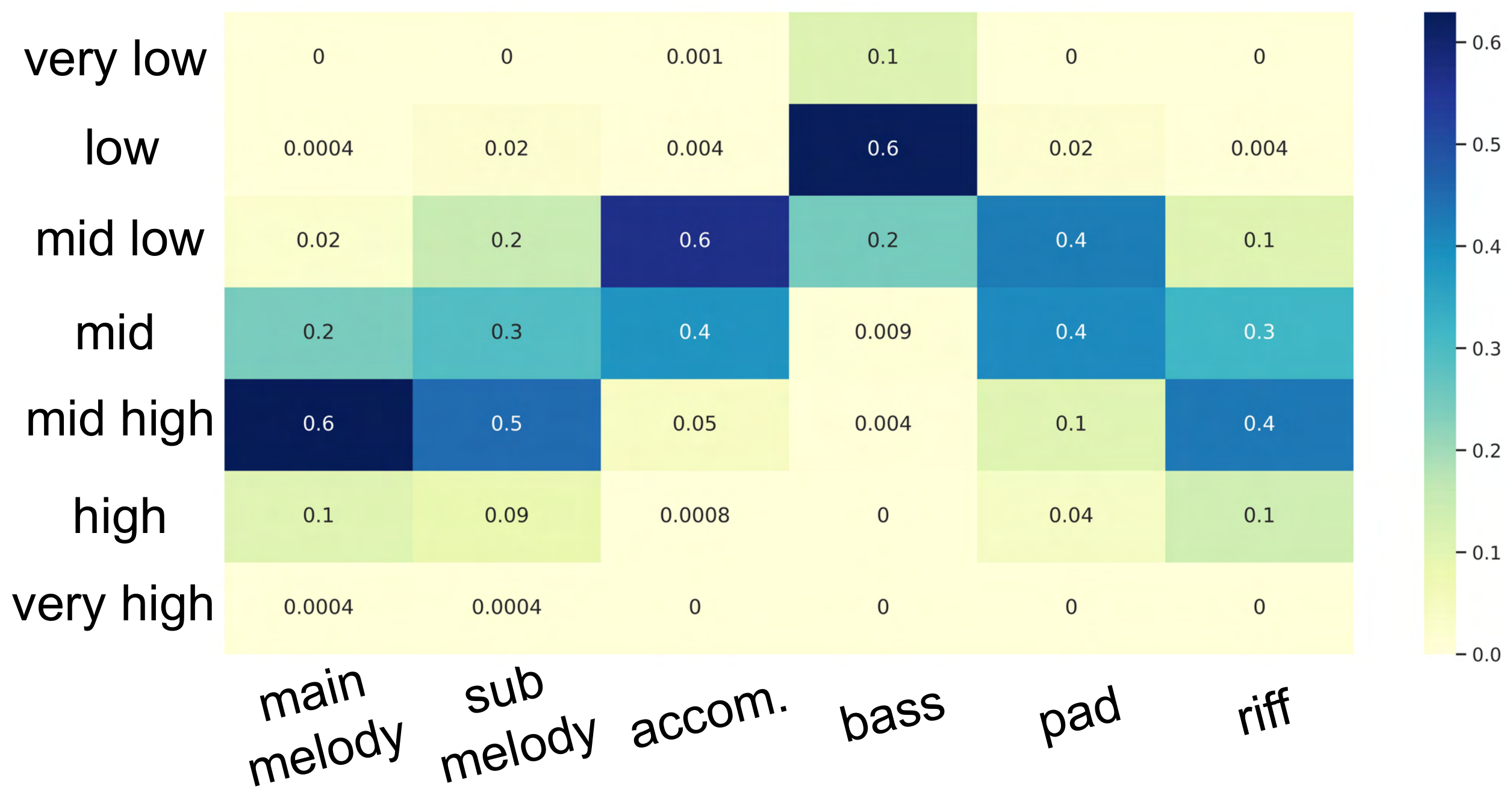}
}
\caption{Correlation heatmap of ComMU metadata. The instrument type and pitch range are different depending on the track-role.}
\label{meta_heatmap}
\vspace{-1mm}
\end{figure*}

\textbf{Analysis between metadata.}
Figure~\ref{meta_heatmap} illustrates the correlation between instruments, track-role, and pitch range. Figure~\ref{meta_heatmap}(a)  shows that keyboards are frequently employed for the main melody and the accompaniment tracks, but pluck strings such as guitars are utilized more for the accompaniment and the riff tracks than for the main melody. In the case of lead instruments, it is rarely used in certain track roles. It can be seen that there is a significant correlation between track-role and instrument. Therefore, when music is generated under the condition of a given instrument without the track role information, the generated sequence is highly likely to be correlated with the track-role. This makes it difficult to make music with a low correlation track-role and musical instrument combination. For example, music with a keyboard as a base and a guitar as a main melody is rarely produced without explicit track-role condition. We conduct an experiment to show this in Section~\ref{5.2_track-role}.

Through the correlation between the track-role and the pitch range, Figure~\ref{meta_heatmap}(b) demonstrates that the pitch range primarily used varies depending on the track-role. The pitch of the melody track is relatively high, while the accompaniment and pad tracks are distributed one level lower. The bass track has the lowest range of notes, following the common sense, and the riff track has a high pitch range similar to the sub melody. This is because the sub melody and riff tracks have the same role in supporting the main melody and mainly use the pitch range that the main instrument does not use. See Appendix~\ref{appendix:Additional_Data_Analysis} for more details about data analysis and limitations.


\section{Experiments}
    In this section, we examine \emph{combinatorial music generation} with the ComMU dataset. For evaluation, we set 90\% of the data for training and hold out the rest for validation. All the metadata in the validation set is exclusive to the training set. We apply the data augmentation method defined in Appendix~\ref{appendix:A_1} to the training set.

\subsection{Problem definition} \label{problem_def}
Our task aims to generate music with specific musical metadata. Given a ComMU sample (Appendix, Figure~\ref{fig:encode}) $X=\{x_{1}^M,.., x_{11}^{M}, x_{12}^{S},.., x_{N}^S\}$, where $M$ and $S$ indicate the tokens ($x_n$) for the metadata and the note sequence respectively, and $x_{N}^S$ is the $\mathtt{eos}$ token, we train an auto-regressive language model by maximizing the following log-likelihood function.

\begin{equation}
\label{eq:1}
\mathcal{L}_{\theta}(X) = \sum_{t=12}^{T}\log p_{\theta}(x_t^S \mid x_{< t}).
\end{equation}

When the training is completed, we can generate diverse note sequences with a specific metadata $x^{M}_{1:11}$ and the chord progression  $\mathcal{C} =  \begin{Bmatrix}x_{i}^{chord}\end{Bmatrix}_{i=1}^K \in x^S$ as follows:
\begin{equation}
\label{eq:1}
\widehat{x_t^S} = g(p_{\theta}(x_t^S \mid x_{1:11}^M, \mathcal{C})), 
\end{equation}
where K is the length of the chord progression, and $g$ is a decoding algorithm. We use Top-k~\cite{fan2018hierarchical, holtzman2018learning} sampling for the decoding.

\begin{figure}
\centering
\includegraphics[width=0.67\textwidth, height=45mm]{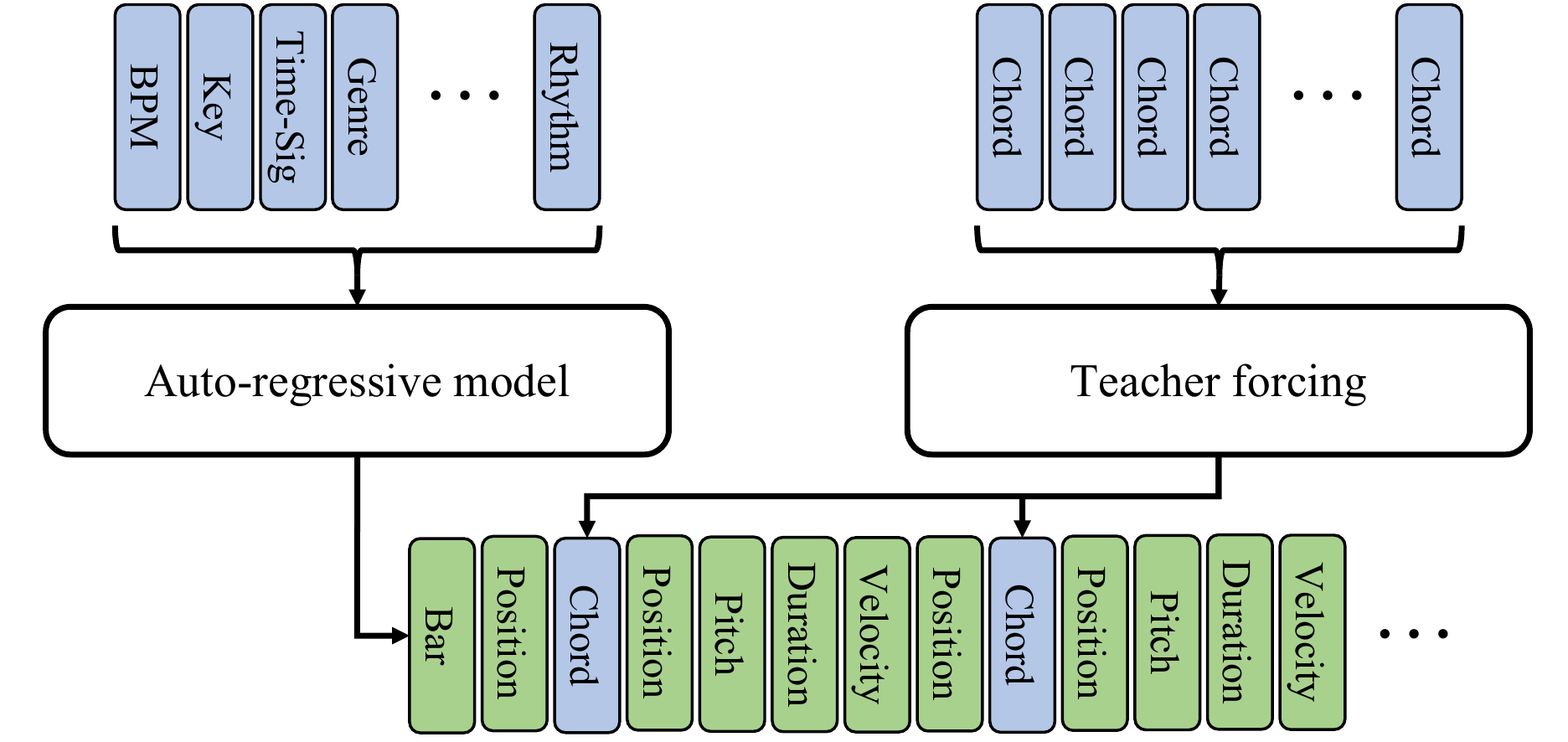}
\caption{\label{fig:inference}Architecture for the $\mathtt{stage 1}$ of combinatorial music generation at inference phase.}
\end{figure}

In music generation, the auto-regressive language model with the transformer structure has shown to be very effective~\cite{huang2018music, payne2019musenet, jiang2020transformer, muhamed2021symbolic}. Among them, we use Transformer-XL~\cite{dai2019transformer} as the baseline to illustrate the effectiveness of ComMU on $\mathtt{stage 1}$ of combinatorial music generation.
During the inference phase, we insert chord tokens in order to infuse chord progression (Figure~\ref{fig:inference}).

\subsection{Evaluation metric}

It is worth mentioning that the objective evaluation of generated music quality is still an open issue~\cite{wu2020jazz}. However, the goal of our task, which is to generate various high-quality music while being well controlled by metadata, is clear. In line with these goals, we evaluate the generated music on three criteria: controllability, diversity, and fidelity. We generate samples one by one from the validation metadata to measure the controllability. For measuring diversity, we generate 10 samples per one validation metadata.

\textbf{Controllability.} We evaluate how accurately the given metadata controls the generated music. We focus on pitch, velocity, and harmony(key, chord), which are clearly measurable among the 12 metadata. The controllability of the pitch is measured by the ratio of notes that meets the given pitch range among all generated note sequences. The number of notes within the min/max velocity range is used to measure the controllability of velocity. For evaluating harmonies, we check whether the pitch of note is within the scale of the corresponding audio key. If the pitch is out of scale, we check whether the pitch matches the chord tone over the duration.
If both conditions are not satisfied, the note is evaluated as dissonant. 
The number of notes evaluated as not dissonant is used as the metric to quantify the controllability of the harmony.

\textbf{Diversity.} We define diversity metric as the average pairwise distance between multiple music generated from the same metadata. Although there is no general way to measure the distance between music, it can be defined by using chroma~\cite{fujishima1999real} and groove similarities~\cite{dixon2004towards}, which measure the cosine similarity of pitch class and rhythm between two music~\cite{wu2021musemorphose}. We define the distance and the diversity metric as follows:

\begin{equation}
\label{eq:3}
\;\; \mathtt{dist(o_i, o_j)} = \sqrt{\frac{(1-\mathtt{sim_{chr}}(o_i, o_j))^{2}+(1-\mathtt{sim_{grv}}(o_i, o_j))^{2}}{2}},
\end{equation}
\begin{equation}
\label{eq:4}
\;\; \mathtt{diversity(\mathcal{O})} =  \frac{1}{\binom{n}{2}}\sum_{i=1}^{n}\sum_{j=1}^{n} \mathtt{dist}(o_{i}, o_{j})\,\,,\,\,  i < j,
\end{equation}
where $\mathcal{O}= \begin{Bmatrix}o_{i}\end{Bmatrix}_{i=1}^n$, $\mathtt{sim_{chr}}$, and $\mathtt{sim_{grv}}$ denote respectively $n$ piece of generated music with same metadata, chroma similarity and groove similarity.

\textbf{Fidelity.} We define the fidelity as win rates of generated samples against real samples. To this end, we have conducted a survey where participants were asked to select their preference between the generated and the real sample with the same metadata.




\subsection{Results} 
\label{results}
\begin{table}
\centering
\begin{tabular}{c@{}c}

\resizebox{0.5\linewidth}{!}{%
\begin{tabular}{l|rrr|r}
  \toprule
    &\multicolumn{3}{c|}{Controllability} & Diversity \\
    \multicolumn{1}{c|}{K, $\tau$} & \multicolumn{1}{c}{CP $\uparrow$} & \multicolumn{1}{c}{CV $\uparrow$} & \multicolumn{1}{c|}{CH $\uparrow$} & \multicolumn{1}{c}{D $\uparrow$} \\
  \midrule
32, 0.7   & 0.8798 & 0.9696 & 0.9976 & 0.2626  \\
32, 0.95  & 0.8412 & 0.9102 & 0.9946 & 0.3160  \\
32, 1.2   & 0.7721 & 0.8566 & 0.9910 & 0.3688  \\
100, 0.95 & 0.8609 & 0.9138 & 0.9954 & 0.3195  \\
200, 0.95 & 0.8488 & 0.9043 & 0.9958 & 0.3165  \\
    \bottomrule
  \end{tabular}}&\hspace{1mm} 
\raisebox{-.475\height}{\includegraphics[width=0.47\linewidth]{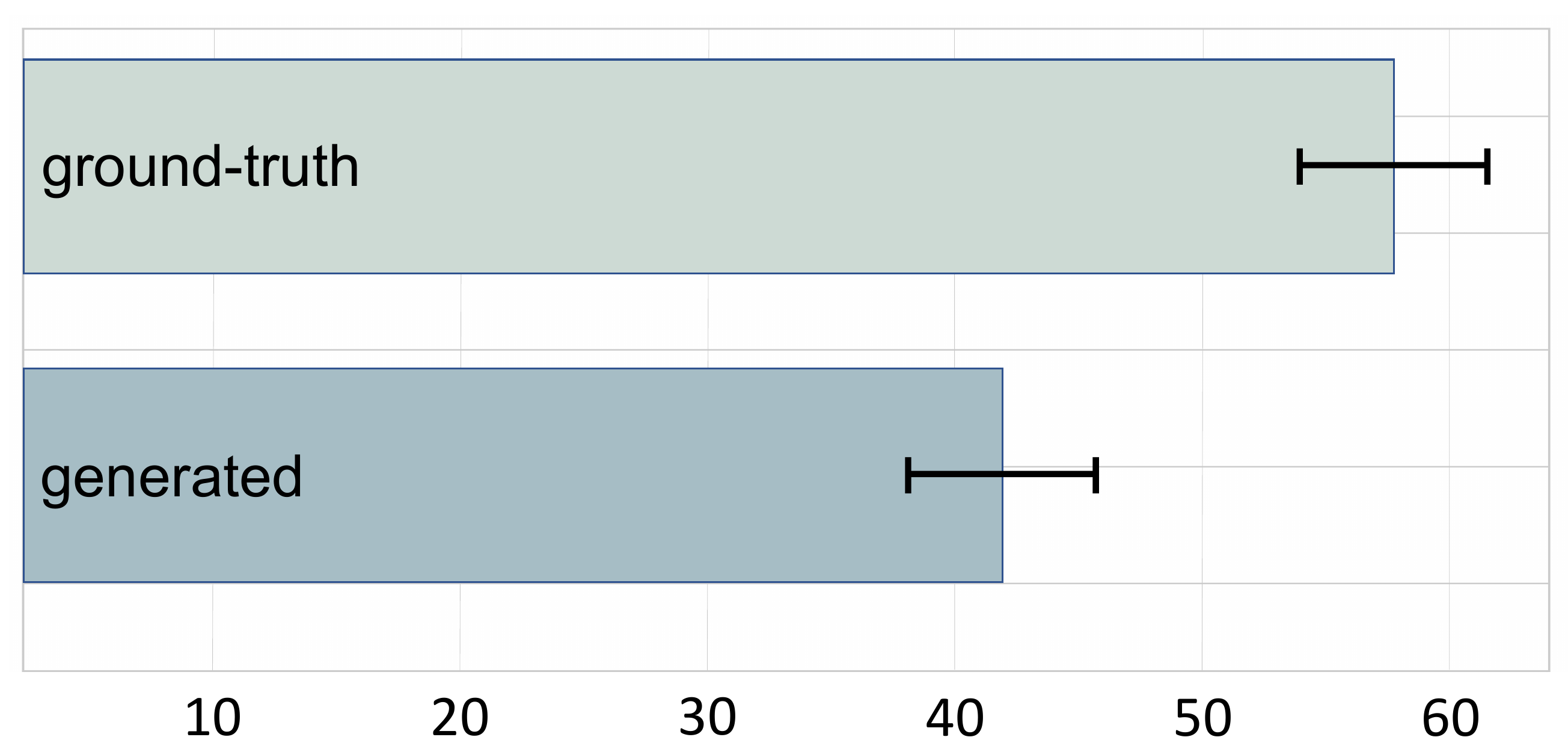}}\vspace{1.5mm}
\\
\small (a) Objective metric result &\hspace{1mm} \small (b) Win rates of generated vs. ground-truth(\%)
\end{tabular}
\vspace{2mm}
\caption{Evaluation results. Objective metric include pitch control (CP), velocity control (CV), harmony control (CH), and diversity (D). We measure win rates using generated samples with K: 32, $\tau$: 0.95.}
\label{result_table}
\vspace{-3mm}
\end{table}



In the language model, the sampling parameter top-k ($K$) and the softmax temperature ($\tau$)~\cite{ackley1985learning} affect the generated sequences. We explore the controllability and the diversity of generated music by adjusting $K$ and $\tau$. Table~\ref{result_table}(a) shows that the controllability declines and the diversity rises as $\tau$ increases. However, the parameter change does not affect the controllability of the harmony because the explicit chord progression induces accurate musical coherence. Moreover, unlike text generation, even if K increases, the change in the results is negligible. This is because the vocabulary size of music (i.e., ComMU) is way smaller than a general language model. Empirical results reveal discrepancies between language and music, and we hope that research on model architecture suitable for music generation will be conducted in the future.

For evaluating fidelity, we ask 40 anonymous composers to 30 questions comparing generated samples to ground-truth samples with the same metadata through Amazon Mechanical Turk. Table~\ref{result_table}(b) shows the average of win rates of generated samples against real samples. This shows how close the generated music is to the human level. It does not beat the ground-truth samples yet, but it shows a fairly close performance. See Appendix~\ref{appendix:amazon_mturk} for details about Amazon Mechanical Turk. Qualitative results can be found in \url{https://pozalabs.github.io/ComMU/}.

\section{Discussion}
    \begin{figure*}[!t]
\centering
\subfigure[Experiment with sus4 token]{
\includegraphics[width=0.48\columnwidth, height=40mm]{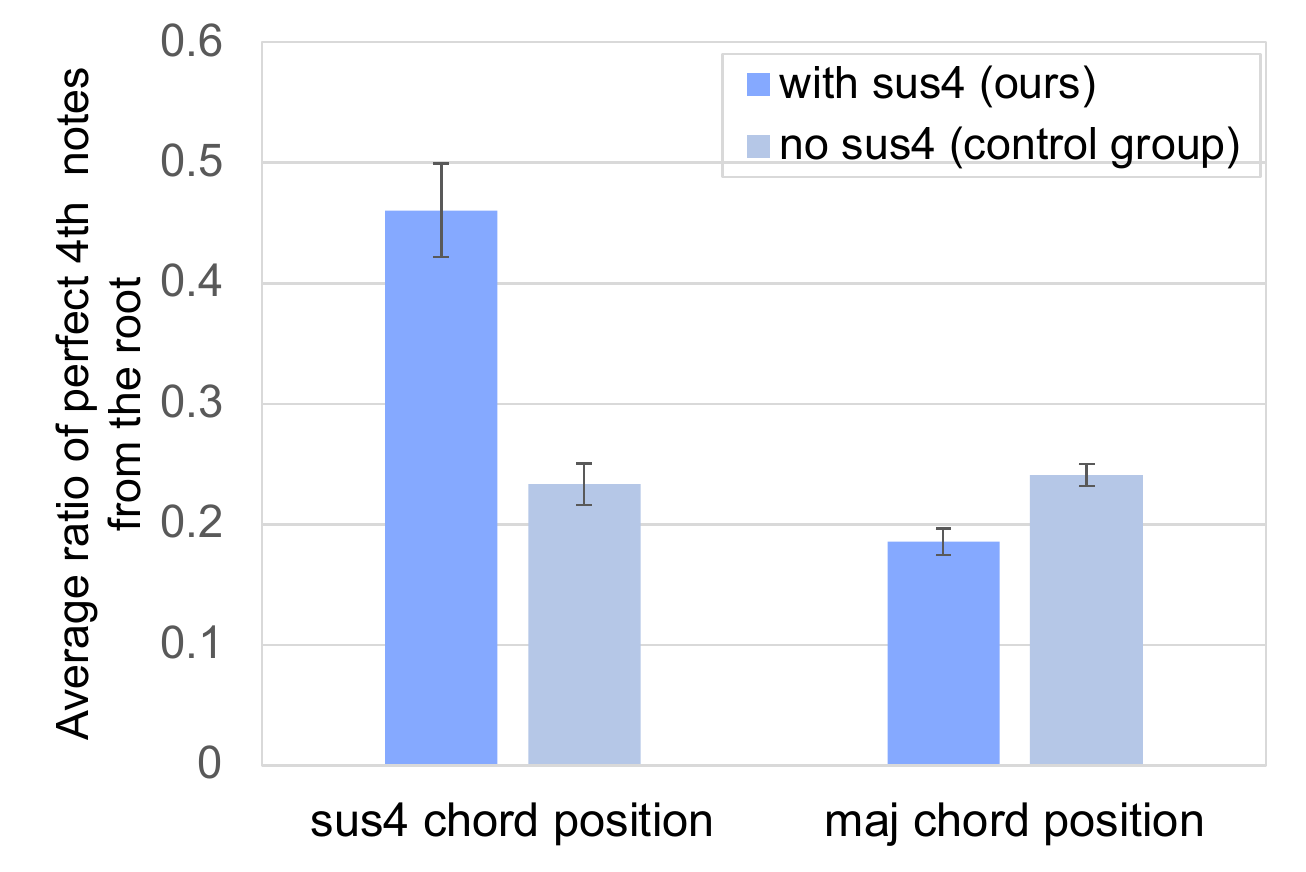}
\label{fig:chord_quality_a}
}
\subfigure[Experiment with maj7 token]{
\includegraphics[width=0.48\columnwidth, height=40mm]{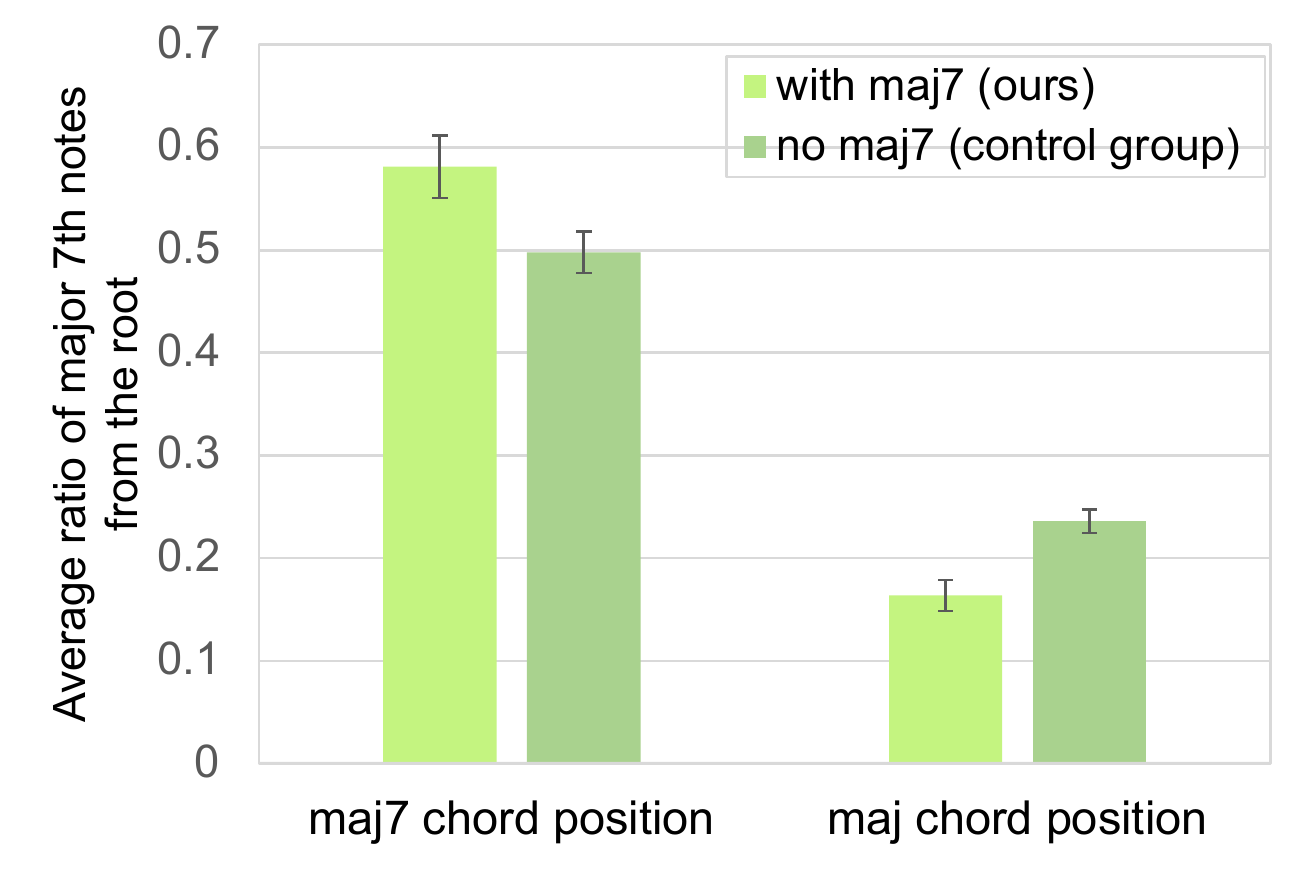}
\label{fig:chord_quality_b}
}
\caption{\textbf{Ablation study results of chord quality.} We take sus4 and maj7 chord quality for all chord roots, respectively. Sus4 is a chord quality in which the major 3rd is replaced with a perfect 4th. Maj7 is consisted of the root, major 3rd, perfect 5th, and major 7th. Therefore, compared to the major chord (root, major 3rd, perfect 5th), the sus4 can produce perfect 4th notes, and the maj7 can produce major 7th notes.}
\label{fig:chord_quality}
\vspace{-2mm}
\end{figure*}

Our rich metadata makes it easy to reflect the composer's intention and increases the capacity and the flexibility of the automatic composition. Since track-role and extended chord quality are primary metadata that distinguishes ComMU from other datasets, we demonstrate the advantage of unique metadata through ablation studies. 

\subsection{Extended chord quality}

Chord quality of ComMU not only has major(maj), minor(min), diminished, augmented, and dominant but also incorporates sus4, maj7, half-diminished, and min7. Extended chord quality makes it easier to understand the precise intention of composition and leads to the improvement in harmonic performance. We experiment with how the extended chord quality affects the note creation. 

In this experiment, we compare the note sequence produced by two different models where the experimental group (ours) holds sus4 or maj7 tokens while the control group does not. The chord ground truth depends on whether the composer initially intended to write a sus4 or maj7 chord in a particular position in the sequence. Here we compare the average ratio of perfect 4th and major 7th notes present in those positions depending on their chord ground truth. In other words, the right halves of both Figure~\ref{fig:chord_quality_a} and ~\ref{fig:chord_quality_b} refer to the chord ground truth of the major chord, whereas the left halves of both ~\ref{fig:chord_quality_a} and ~\ref{fig:chord_quality_b} refer to the chord ground truth of sus4 and maj7 chord respectively.

As shown in Figure~\ref{fig:chord_quality_a}, the experimental group better adheres to the sus4 chord scale, where the average ratio of perfect 4th notes produced is more than 0.2 higher in the experimental than in the control group. In the case of major chord position, the experimental group produced less perfect 4th notes than the control group, better conforming to the major chord scale than its counterpart do. Such is also true in the case of major7 chords in Figure~\ref{fig:chord_quality_b}. This demonstrates that whether the composer intended to write major or extended chords (sus4 and maj7) in a particular position in the sequence, the model trained with the extended chord tokens better reflects his or her intention.

\subsection{Multi-track with track-role} \label{5.2_track-role}
\begin{figure*}[!t]
\centering
\subfigure[Response distribution of piano samples]{
\includegraphics[width=0.48\textwidth, height=38mm]{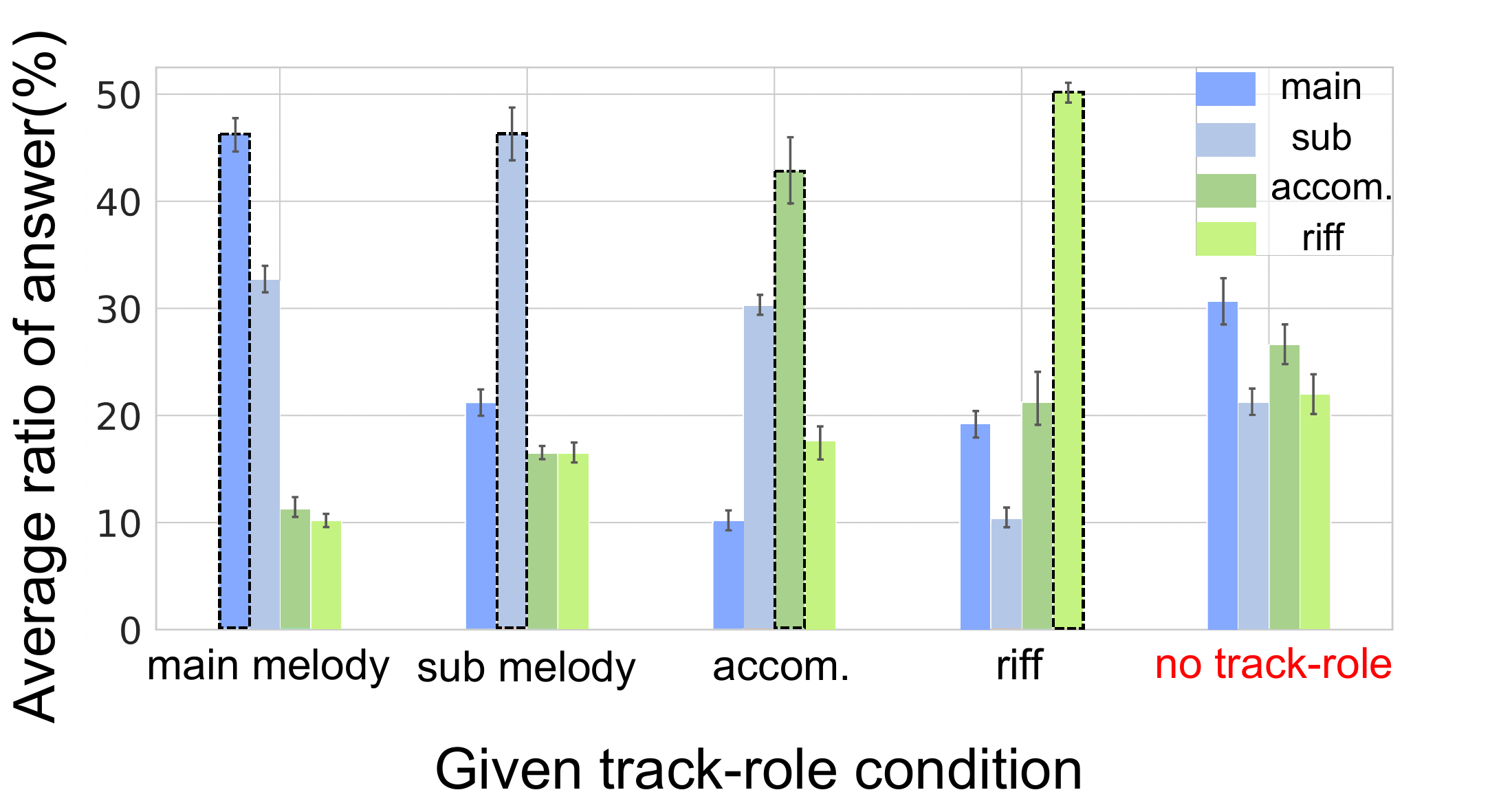}
}
\subfigure[Response distribution of string samples]{
\includegraphics[width=0.48\textwidth, height=38mm]{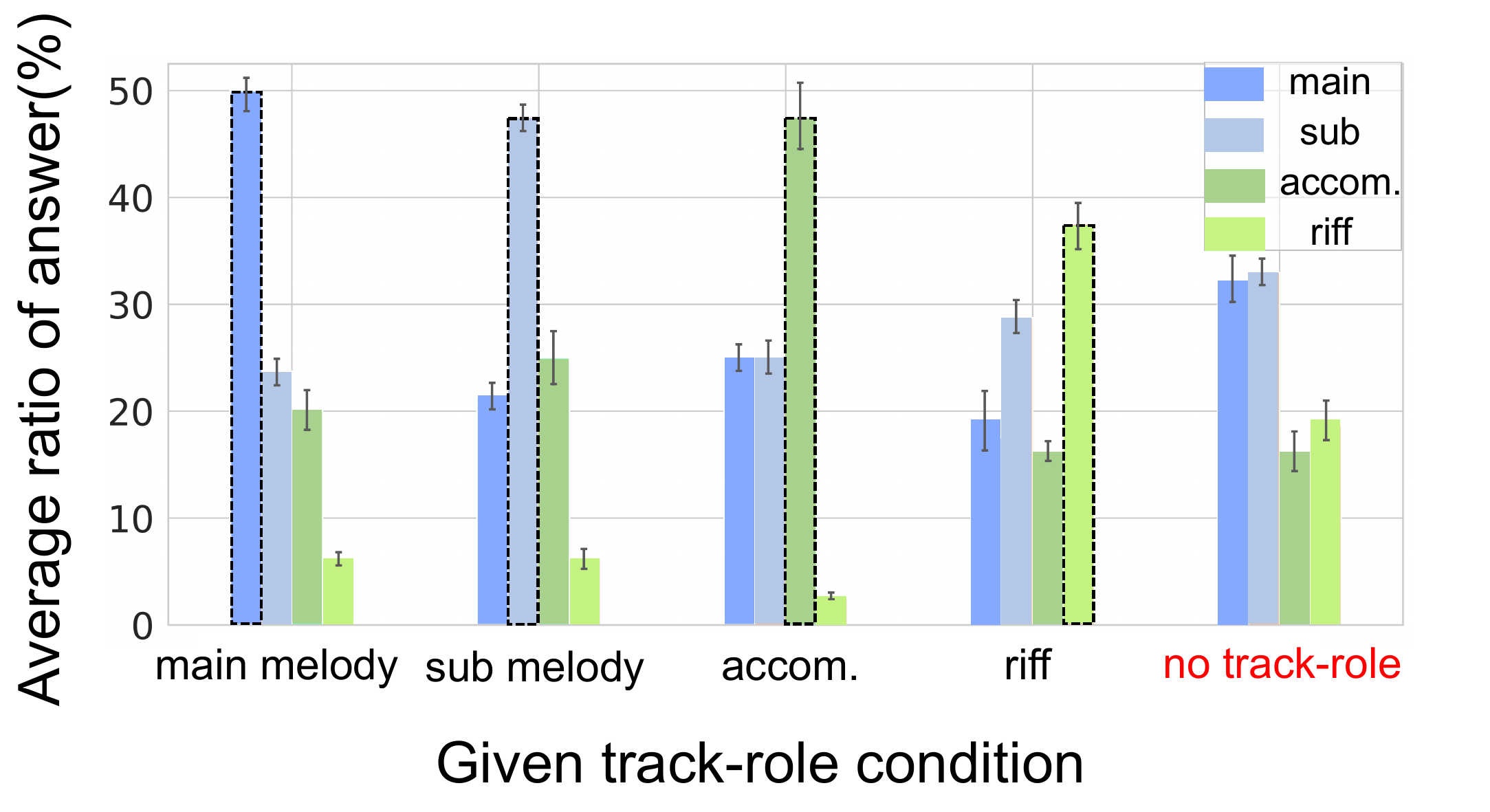}
}
\caption{Response of track-role classification survey. It illustrates that appropriate music is created when the role of each track is specified as a condition.}
\label{fig:mturk_track-role}
\vspace{-1mm}
\end{figure*}

While early literature has configured multi-track with instruments~\cite{ens2020mmm, dong2018musegan}, ComMU introduces track-role to configure multi-track. This can improve the capacity and flexibility of automatic composition and give more detailed conditions for music generation. For example, we can make music consisting of piano and guitar without track-role information. However, we cannot make music with piano as an accompaniment and guitar as a melody track. This is because even the same instrument has a different note shape, pitch range, and rhythm depending on the track-role(Figure~\ref{meta_heatmap}).

To demonstrate the impact of the track-role, we conduct experiments comparing generated music, including track-role, with music that does not. We play 64 generated music samples through Amazon Mechanical Turk to 20 anonymous professional composers worldwide and ask which track-role is most appropriate for each music.

According to the survey result in Figure~\ref{fig:mturk_track-role}, if no track-role information is provided, the response resembles the track-role distribution by the instrument depicted in Figure~\ref{meta_heatmap}. This indicates that the track-role is generated at random. Conversely, when a track-role is given, many subjects respond that the given track-role is the most appropriate. This indicates that our track-role metadata provides significant guidance to the generated music.
\section{Conclusion} \label{conclusion}
    In this paper, we attempted to push the boundaries of automatic composition by introducing combinatorial music generation.
We presented ComMU, a dataset for combinatorial music generation $\mathtt{stage 1}$, consisting of 12 metadata matched with note sequences manually created by composers.
Furthermore, we present quantitative evaluations for generated music through fidelity, controllability, and diversity, and we demonstrate the beneﬁts of unique metadata such as track-role and extended chord quality. We still need the generated note sequences to be combined by experts in $\mathtt{stage 2}$, but by automating $\mathtt{stage 1}$, we have dramatically reduced the time it takes to compose music. 
With the development of $\mathtt{stage 2}$ in the future, we expect the automatic composition to be close to the human level. It is worthwhile to notice that combinatorial music generation would be one of the potential uses of the ComMU dataset. We hope that ComMU opens up a wide range of future research on automatic composition.

\textbf{Acknowledgment.} This research was supported by Culture, Sports and Tourism R\&D Program through the Korea Creative Content Agency grant funded by the Ministry of Culture, Sports and Tourism in 2022 (Project Name: AI Producer: Developing technology of custom music composition, Project Number: R2022020066, Contribution Rate: 100\%).

\small
\bibliography{neurips_data_2022}

\begin{thebibliography}{43}
\providecommand{\natexlab}[1]{#1}
\providecommand{\url}[1]{\texttt{#1}}
\expandafter\ifx\csname urlstyle\endcsname\relax
  \providecommand{\doi}[1]{doi: #1}\else
  \providecommand{\doi}{doi: \begingroup \urlstyle{rm}\Url}\fi

\bibitem[Nierhaus(2009)]{nierhaus2009historical}
Gerhard Nierhaus.
\newblock Historical development of algorithmic procedures.
\newblock \emph{Algorithmic Composition: Paradigms of Automated Music
  Generation}, pages 7--66, 2009.

\bibitem[Giraudo(2021)]{giraudo2021generation}
Samuele Giraudo.
\newblock Generation of musical patterns through operads.
\newblock \emph{arXiv preprint arXiv:2104.12432}, 2021.

\bibitem[Bretan et~al.(2016)Bretan, Weinberg, and Heck]{bretan2016}
Mason Bretan, Gil Weinberg, and Larry Heck.
\newblock A unit selection methodology for music generation using deep neural
  networks.
\newblock In \emph{International Conference on Computational Creativity
  (ICCC)}, 2016.

\bibitem[Jaques et~al.(2016)Jaques, Gu, Turner, and Eck]{45871}
Natasha Jaques, Shixiang Gu, Richard~E. Turner, and Douglas Eck.
\newblock Generating music by fine-tuning recurrent neural networks with
  reinforcement learning.
\newblock In \emph{Deep Reinforcement Learning Workshop, NIPS}, 2016.

\bibitem[Huang et~al.(2019)Huang, Vaswani, Uszkoreit, Shazeer, Simon,
  Hawthorne, Dai, Hoffman, Dinculescu, and Eck]{huang2018music}
Cheng-Zhi~Anna Huang, Ashish Vaswani, Jakob Uszkoreit, Noam Shazeer, Ian Simon,
  Curtis Hawthorne, Andrew~M Dai, Matthew~D Hoffman, Monica Dinculescu, and
  Douglas Eck.
\newblock Music transformer.
\newblock In \emph{International Conference on Learning Representations
  (ICLR)}, 2019.

\bibitem[Payne(2019)]{payne2019musenet}
Christine Payne.
\newblock Musenet.
\newblock \emph{OpenAI Blog}, 3, 2019.

\bibitem[Jiang et~al.(2020)Jiang, Xia, Carlton, Anderson, and
  Miyakawa]{jiang2020transformer}
Junyan Jiang, Gus~G Xia, Dave~B Carlton, Chris~N Anderson, and Ryan~H Miyakawa.
\newblock Transformer vae: A hierarchical model for structure-aware and
  interpretable music representation learning.
\newblock In \emph{ICASSP 2020-2020 IEEE International Conference on Acoustics,
  Speech and Signal Processing (ICASSP)}, pages 516--520. IEEE, 2020.

\bibitem[Muhamed et~al.(2021)Muhamed, Li, Shi, Yaddanapudi, Chi, Jackson,
  Suresh, Lipton, and Smola]{muhamed2021symbolic}
Aashiq Muhamed, Liang Li, Xingjian Shi, Suri Yaddanapudi, Wayne Chi, Dylan
  Jackson, Rahul Suresh, Zachary~C Lipton, and Alex~J Smola.
\newblock Symbolic music generation with transformer-gans.
\newblock In \emph{Proceedings of the AAAI Conference on Artificial
  Intelligence}, volume~35, pages 408--417, 2021.

\bibitem[Sadie and Tyrrell(2001)]{sadie2001dictionary}
Stanley Sadie and John Tyrrell.
\newblock \emph{Dictionary of music and musicians}.
\newblock New York: Oxford University Press. Y{\'o}natan S{\'a}nchez, 2001.

\bibitem[Raffel(2016)]{raffel2016learning}
Colin Raffel.
\newblock \emph{Learning-based methods for comparing sequences, with
  applications to audio-to-midi alignment and matching}.
\newblock Columbia University, 2016.

\bibitem[Donahue et~al.(2019)Donahue, Mao, Li, Cottrell, and
  McAuley]{donahue2019lakhnes}
Chris Donahue, Huanru~Henry Mao, Yiting~Ethan Li, Garrison~W Cottrell, and
  Julian McAuley.
\newblock Lakhnes: Improving multi-instrumental music generation with
  cross-domain pre-training.
\newblock In \emph{International Society for Music Information Retrieval
  (ISMIR)}, 2019.

\bibitem[Hawthorne et~al.(2019)Hawthorne, Stasyuk, Roberts, Simon, Huang,
  Dieleman, Elsen, Engel, and Eck]{hawthorne2018enabling}
Curtis Hawthorne, Andriy Stasyuk, Adam Roberts, Ian Simon, Cheng-Zhi~Anna
  Huang, Sander Dieleman, Erich Elsen, Jesse Engel, and Douglas Eck.
\newblock Enabling factorized piano music modeling and generation with the
  maestro dataset.
\newblock In \emph{International Conference on Learning Representations
  (ICLR)}, 2019.

\bibitem[Dorfer et~al.(2018)Dorfer, Haji{\v{c}}~Jr, Arzt, Frostel, and
  Widmer]{dorfer2018learning}
Matthias Dorfer, Jan Haji{\v{c}}~Jr, Andreas Arzt, Harald Frostel, and Gerhard
  Widmer.
\newblock Learning audio--sheet music correspondences for cross-modal retrieval
  and piece identification.
\newblock \emph{Transactions of the International Society for Music Information
  Retrieval}, 1\penalty0 (1), 2018.

\bibitem[Ferreira et~al.(2020)Ferreira, Lelis, and
  Whitehead]{ferreira2020computer}
Lucas Ferreira, Levi Lelis, and Jim Whitehead.
\newblock Computer-generated music for tabletop role-playing games.
\newblock In \emph{Proceedings of the AAAI Conference on Artificial
  Intelligence and Interactive Digital Entertainment}, volume~16, pages 59--65,
  2020.

\bibitem[Kong et~al.(2022)Kong, Li, Chen, and Wang]{kong2020giantmidi}
Qiuqiang Kong, Bochen Li, Jitong Chen, and Yuxuan Wang.
\newblock Giantmidi-piano: A large-scale midi dataset for classical piano
  music.
\newblock \emph{Transactions of the International Society for Music Information
  Retrieval}, 5\penalty0 (1):\penalty0 87--98, 2022.

\bibitem[Shih et~al.(2022)Shih, Wu, Zalkow, Muller, and Yang]{shih2022theme}
Yi-Jen Shih, Shih-Lun Wu, Frank Zalkow, Meinard Muller, and Yi-Hsuan Yang.
\newblock Theme transformer: Symbolic music generation with theme-conditioned
  transformer.
\newblock \emph{IEEE Transactions on Multimedia}, 2022.

\bibitem[von R{\"u}tte et~al.(2022)von R{\"u}tte, Biggio, Kilcher, and
  Hoffman]{von2022figaro}
Dimitri von R{\"u}tte, Luca Biggio, Yannic Kilcher, and Thomas Hoffman.
\newblock Figaro: Generating symbolic music with fine-grained artistic control.
\newblock \emph{arXiv preprint arXiv:2201.10936}, 2022.

\bibitem[Wu and Yang(2021)]{wu2021musemorphose}
Shih-Lun Wu and Yi-Hsuan Yang.
\newblock Muse{M}orphose: Full-song and fine-grained music style transfer with
  just one {T}ransformer {VAE}.
\newblock \emph{arXiv preprint arXiv:2105.04090}, 2021.

\bibitem[Ens and Pasquier(2020)]{ens2020mmm}
Jeff Ens and Philippe Pasquier.
\newblock Mmm: Exploring conditional multi-track music generation with the
  transformer.
\newblock \emph{arXiv preprint arXiv:2008.06048}, 2020.

\bibitem[Dong et~al.(2018)Dong, Hsiao, Yang, and Yang]{dong2018musegan}
Hao-Wen Dong, Wen-Yi Hsiao, Li-Chia Yang, and Yi-Hsuan Yang.
\newblock Musegan: Multi-track sequential generative adversarial networks for
  symbolic music generation and accompaniment.
\newblock In \emph{Thirty-Second AAAI Conference on Artificial Intelligence},
  2018.

\bibitem[Zhu et~al.(2018)Zhu, Liu, Yuan, Qin, Li, Zhang, Zhou, Wei, Xu, and
  Chen]{zhu2018xiaoice}
Hongyuan Zhu, Qi~Liu, Nicholas~Jing Yuan, Chuan Qin, Jiawei Li, Kun Zhang,
  Guang Zhou, Furu Wei, Yuanchun Xu, and Enhong Chen.
\newblock Xiaoice band: A melody and arrangement generation framework for pop
  music.
\newblock In \emph{Proceedings of the 24th ACM SIGKDD International Conference
  on Knowledge Discovery \& Data Mining}, pages 2837--2846, 2018.

\bibitem[Zhu et~al.(2020)Zhu, Liu, Yuan, Zhang, Zhou, and Chen]{zhu2020pop}
Hongyuan Zhu, Qi~Liu, Nicholas~Jing Yuan, Kun Zhang, Guang Zhou, and Enhong
  Chen.
\newblock Pop music generation: From melody to multi-style arrangement.
\newblock \emph{ACM Transactions on Knowledge Discovery from Data (TKDD)},
  14\penalty0 (5):\penalty0 1--31, 2020.

\bibitem[Hung et~al.(2021)Hung, Ching, Doh, Kim, Nam, and Yang]{hung2021emopia}
Hsiao-Tzu Hung, Joann Ching, Seungheon Doh, Nabin Kim, Juhan Nam, and Yi-Hsuan
  Yang.
\newblock Emopia: A multi-modal pop piano dataset for emotion recognition and
  emotion-based music generation.
\newblock In \emph{International Society for Music Information Retrieval
  Conference (ISMIR)}, 2021.

\bibitem[Di et~al.(2021)Di, Jiang, Liu, Wang, Zhu, He, Liu, and
  Yan]{di2021video}
Shangzhe Di, Zeren Jiang, Si~Liu, Zhaokai Wang, Leyan Zhu, Zexin He, Hongming
  Liu, and Shuicheng Yan.
\newblock Video background music generation with controllable music
  transformer.
\newblock In \emph{Proceedings of the 29th ACM International Conference on
  Multimedia}, pages 2037--2045, 2021.

\bibitem[Yang et~al.(2017)Yang, Chou, and Yang]{Yang2017MidiNetAC}
Li-Chia Yang, Szu-Yu Chou, and Yi-Hsuan Yang.
\newblock Midinet: A convolutional generative adversarial network for
  symbolic-domain music generation.
\newblock In \emph{International Society for Music Information Retrieval
  Conference (ISMIR)}, 2017.

\bibitem[Tan(2019)]{tan2019chordal}
Hao~Hao Tan.
\newblock Chordal: A chord-based approach for music generation using bi-lstms.
\newblock In \emph{International Conference on Computational Creativity
  (ICCC)}, pages 364--365, 2019.

\bibitem[Hakimi et~al.(2020)Hakimi, Bhonker, and
  El-Yaniv]{Hakimi2020BebopNetDN}
Shunit~Haviv Hakimi, Nadav Bhonker, and Ran El-Yaniv.
\newblock Bebopnet: Deep neural models for personalized jazz improvisations.
\newblock In \emph{International Society for Music Information Retrieval
  (ISMIR)}, 2020.

\bibitem[Liutkus et~al.(2017)Liutkus, St{\"o}ter, Rafii, Kitamura, Rivet, Ito,
  Ono, and Fontecave]{SiSEC16}
Antoine Liutkus, Fabian-Robert St{\"o}ter, Zafar Rafii, Daichi Kitamura,
  Bertrand Rivet, Nobutaka Ito, Nobutaka Ono, and Julie Fontecave.
\newblock The 2016 signal separation evaluation campaign.
\newblock In \emph{International conference on latent variable analysis and
  signal separation}, pages 323--332. Springer, 2017.

\bibitem[Rafii et~al.(2017)Rafii, Liutkus, St{\"o}ter, Mimilakis, and
  Bittner]{rafii2017musdb18}
Zafar Rafii, Antoine Liutkus, Fabian-Robert St{\"o}ter, Stylianos~Ioannis
  Mimilakis, and Rachel Bittner.
\newblock Musdb18-a corpus for music separation.
\newblock 2017.

\bibitem[Huang and Yang(2020)]{huang2020pop}
Yu-Siang Huang and Yi-Hsuan Yang.
\newblock Pop music transformer: Beat-based modeling and generation of
  expressive pop piano compositions.
\newblock In \emph{Proceedings of the 28th ACM International Conference on
  Multimedia}, pages 1180--1188, 2020.

\bibitem[Vaswani et~al.(2017)Vaswani, Shazeer, Parmar, Uszkoreit, Jones, Gomez,
  Kaiser, and Polosukhin]{vaswani2017attention}
Ashish Vaswani, Noam Shazeer, Niki Parmar, Jakob Uszkoreit, Llion Jones,
  Aidan~N Gomez, {\L}ukasz Kaiser, and Illia Polosukhin.
\newblock Attention is all you need.
\newblock In \emph{Advances in Neural Information Processing Systems
  (NeurIPS)}, pages 5998--6008, 2017.

\bibitem[Fan et~al.(2018)Fan, Lewis, and Dauphin]{fan2018hierarchical}
Angela Fan, Mike Lewis, and Yann Dauphin.
\newblock Hierarchical neural story generation.
\newblock In \emph{Proceedings of the 56th Annual Meeting of the Association
  for Computational Linguistics (ACL)}, pages 889--898, 2018.

\bibitem[Holtzman et~al.(2018)Holtzman, Buys, Forbes, Bosselut, Golub, and
  Choi]{holtzman2018learning}
Ari Holtzman, Jan Buys, Maxwell Forbes, Antoine Bosselut, David Golub, and
  Yejin Choi.
\newblock Learning to write with cooperative discriminators.
\newblock In \emph{Annual Meeting of the Association for Computational
  Linguistics (ACL)}, 2018.

\bibitem[Dai et~al.(2019)Dai, Yang, Yang, Carbonell, Le, and
  Salakhutdinov]{dai2019transformer}
Zihang Dai, Zhilin Yang, Yiming Yang, Jaime~G Carbonell, Quoc Le, and Ruslan
  Salakhutdinov.
\newblock Transformer-xl: Attentive language models beyond a fixed-length
  context.
\newblock In \emph{Proceedings of the 57th Annual Meeting of the Association
  for Computational Linguistics (ACL)}, pages 2978--2988, 2019.

\bibitem[Wu and Yang(2020)]{wu2020jazz}
Shih-Lun Wu and Yi-Hsuan Yang.
\newblock The jazz transformer on the front line: Exploring the shortcomings of
  ai-composed music through quantitative measures.
\newblock \emph{International Society for Music Information Retrieval (ISMIR)},
  2020.

\bibitem[Fujishima(1999)]{fujishima1999real}
Takuya Fujishima.
\newblock Real-time chord recognition of musical sound: A system using common
  lisp music.
\newblock In \emph{International Computer Music Conference (ICMC)}, pages
  464--467, 1999.

\bibitem[Dixon et~al.(2004)Dixon, Gouyon, Widmer, et~al.]{dixon2004towards}
Simon Dixon, Fabien Gouyon, Gerhard Widmer, et~al.
\newblock Towards characterisation of music via rhythmic patterns.
\newblock In \emph{International Society for Music Information Retrieval
  (ISMIR)}, 2004.

\bibitem[Ackley et~al.(1985)Ackley, Hinton, and Sejnowski]{ackley1985learning}
David~H Ackley, Geoffrey~E Hinton, and Terrence~J Sejnowski.
\newblock A learning algorithm for boltzmann machines.
\newblock \emph{Cognitive science}, 9\penalty0 (1):\penalty0 147--169, 1985.

\bibitem[Klapuri(2006)]{klapuri2006introduction}
Anssi Klapuri.
\newblock Introduction to music transcription.
\newblock In \emph{Signal processing methods for music transcription}, pages
  3--20. Springer, 2006.

\bibitem[Plack et~al.(2006)Plack, Oxenham, and Fay]{plack2006pitch}
Christopher~J Plack, Andrew~J Oxenham, and Richard~R Fay.
\newblock \emph{Pitch: neural coding and perception}, volume~24.
\newblock Springer Science \& Business Media, 2006.

\bibitem[Lomas and Xue(2022)]{lomas2022harmony}
J~Derek Lomas and Haian Xue.
\newblock Harmony in design: A synthesis of literature from classical
  philosophy, the sciences, economics, and design.
\newblock \emph{She Ji: The Journal of Design, Economics, and Innovation},
  8\penalty0 (1):\penalty0 5--64, 2022.

\bibitem[Benward and Saker(2003)]{benward2003music}
B.~Benward and M.N. Saker.
\newblock \emph{Music in Theory and Practice}.
\newblock Number V. 1 in Music in Theory and Practice. McGraw-Hill, 2003.
\newblock ISBN 9780072942620.

\bibitem[Rothstein(1992)]{rothstein1992midi}
Joseph Rothstein.
\newblock \emph{MIDI: A comprehensive introduction}, volume~7.
\newblock AR Editions, Inc., 1992.

\end{thebibliography}
\bibliographystyle{unsrtnat}

\section*{Checklist}
\begin{enumerate}

\item For all authors...
\begin{enumerate}
  \item Do the main claims made in the abstract and introduction accurately reflect the paper's contributions and scope?
    \answerYes{See Section~\ref{intro}.}
  \item Did you describe the limitations of your work?
    \answerYes{See Section~\ref{conclusion} and appendix~\ref{appendix:Additional_Data_Analysis}.}
  \item Did you discuss any potential negative societal impacts of your work?
    \answerYes{See appendix~\ref{appendix:societal_impact}.}
  \item Have you read the ethics review guidelines and ensured that your paper conforms to them?
    \answerYes{We have read them and confirmed to them.}
\end{enumerate}

\item If you are including theoretical results...
\begin{enumerate}
  \item Did you state the full set of assumptions of all theoretical results?
    \answerNA{}
	\item Did you include complete proofs of all theoretical results?
    \answerNA{}
\end{enumerate}

\item If you ran experiments (e.g. for benchmarks)...
\begin{enumerate}
  \item Did you include the code, data, and instructions needed to reproduce the main experimental results (either in the supplemental material or as a URL)?
    \answerYes{\url{https://pozalabs.github.io/ComMU/}}
  \item Did you specify all the training details (e.g., data splits, hyperparameters, how they were chosen)?
    \answerYes{See appendix~\ref{appendix:training_detail}.}
	\item Did you report error bars (e.g., with respect to the random seed after running experiments multiple times)?
    \answerYes{See Table~\ref{result_table} and Figure~\ref{fig:chord_quality}.}
	\item Did you include the total amount of compute and the type of resources used (e.g., type of GPUs, internal cluster, or cloud provider)?
    \answerYes{See appendix~\ref{appendix:training_detail}.}
\end{enumerate}

\item If you are using existing assets (e.g., code, data, models) or curating/releasing new assets...
\begin{enumerate}
  \item If your work uses existing assets, did you cite the creators?
    \answerYes{See Section~\ref{dataset-representation} and~\ref{problem_def}.}
  \item Did you mention the license of the assets?
    \answerYes{REMI~\cite{huang2020pop} and Transformer-XL~\cite{dai2019transformer} code is licensed under the GNU General Public License v3.0 and Apache License 2.0, free for research purposes.}
  \item Did you include any new assets either in the supplemental material or as a URL?
    \answerYes{\url{https://pozalabs.github.io/ComMU/}}
  \item Did you discuss whether and how consent was obtained from people whose data you're using/curating?
    \answerYes{See appendix~\ref{appendix:obtained_consent}.}
  \item Did you discuss whether the data you are using/curating contains personally identifiable information or offensive content?
    \answerNA{}
\end{enumerate}

\item If you used crowdsourcing or conducted research with human subjects...
\begin{enumerate}
  \item Did you include the full text of instructions given to participants and screenshots, if applicable?
    \answerYes{See appendix~\ref{appendix:amazon_mturk}}
  \item Did you describe any potential participant risks, with links to Institutional Review Board (IRB) approvals, if applicable?
    \answerNA{}
  \item Did you include the estimated hourly wage paid to participants and the total amount spent on participant compensation?
        \answerYes{See appendix~\ref{appendix:amazon_mturk}}
\end{enumerate}
\vspace{2cm}
\end{enumerate}


\appendix

\section{Pre-processing and representation} \label{appendix: preprocessing}





\subsection{Data pre-processing} \label{appendix:A_1}
\begin{wraptable}{r}{0.45\linewidth}
\vspace{-5mm}
\caption{Encoding dictionary of ComMU. First value of each metadata is an unknown token(e.g., \emph{bpm: 560} denote unknown bpm).}
\label{table:encoding dict}
\vskip 0.015in
\begin{center}
\begin{small}
\begin{sc}
\begin{tabular}{l|r}
\toprule
token & encode value\\
\midrule
pad & 0\\
eos & 1\\
bar & 2\\
note pitch & 3-130\\
note velocity & 131-194\\
chord & 195-303\\
note duration & 304-431\\
position & 432-559\\
\midrule
bpm & 560-600\\
key & 601-625\\
time signature & 626-629\\
pitch range & 630-637\\
number of measure & 638-640\\
instrument & 641-649\\
genre & 650-652\\
min velocity & 653-717\\
max velocity & 653-718\\
track-role & 719-725\\
rhythm & 726-728\\
\bottomrule
\end{tabular}
\end{sc}
\end{small}
\end{center}
\end{wraptable}
\textbf{Data Augmentation.} 
To enlarge the data through augmentation, we give variations to the BPM (-10, -5, +0, +5, +10) as well as to the audio key (C, C-sharp, D, D-sharp, E, F, F-sharp, G, G-sharp, A, A-sharp, B). Total of 60 variations are generated from a single raw MIDI file by manipulating the BPM and the audio key. Note duration is modified according to changes in bpm, and note pitch and chord progression are modified according to changes in audio key.

\textbf{Encoding.} \label{appendix:encoding}
Each metadata and note sequence is mapped to an integer value according to an encoding dictionary depicted in Table~\ref{table:encoding dict}. REMI~\cite{huang2020pop} code is extended for encoding ComMU.

\textbf{Data Scalability.} 
In addition to ComMU, we are also able to utilize other MIDI datasets. A common MIDI file contains tens to hundreds of bars with multiple tracks. However, we are able to process such files using the steps of slice, chunk, and parse. 
\begin{itemize}
\item slice: segmenting the parts where audio key or time signature change occur.
\item chunk: separating tracks and segmenting sections which notes mainly appear as one chunk, and discarding sections where notes do not appear.
\item parse: splitting a MIDI file into certain number of measures (4, 8, 16) to better fit with ComMU.
\end{itemize}

\subsection{Data representation}

\par
ComMU extends REMI by adding 11 metadata excluding chord, which is already present in REMI~\cite{huang2020pop}. REMI includes a $\mathtt{velocity}$ token that denotes velocity, a $\mathtt{pitch}$ token that indicates the beginning of a note with a specific pitch, and a $\mathtt{duration}$ token that measures the length of a note. Further, a $\mathtt{bar}$ token denotes the start of a bar, a $\mathtt{position}$ token indicates the note or chord position in the bar, and $\mathtt{tempo}$ and $\mathtt{chord}$ token represent a tempo change and chord information respectively. Differences between REMI and ComMU are further discussed below.

First, the most significant difference between ComMU and REMI is that ComMU eliminates the tempo token. Although change in tempo often occurs in a complete music, it rarely occurs in short note sequences. Hence, we consider it to be non-essential for generating short music samples. Instead, we fix the tempo of each sample and then express the difference between samples through BPM.

Another difference is that we add 4 extended chord qualities of major 7th, minor 7th, half-diminished(m7b5), and sus4 on top of the existing chord qualities(major, minor, dominant, diminished, and augmented) in REMI. The total number of chord events in ComMU representation is 108.

\begin{figure}
\centering
\includegraphics[width=0.95\textwidth]{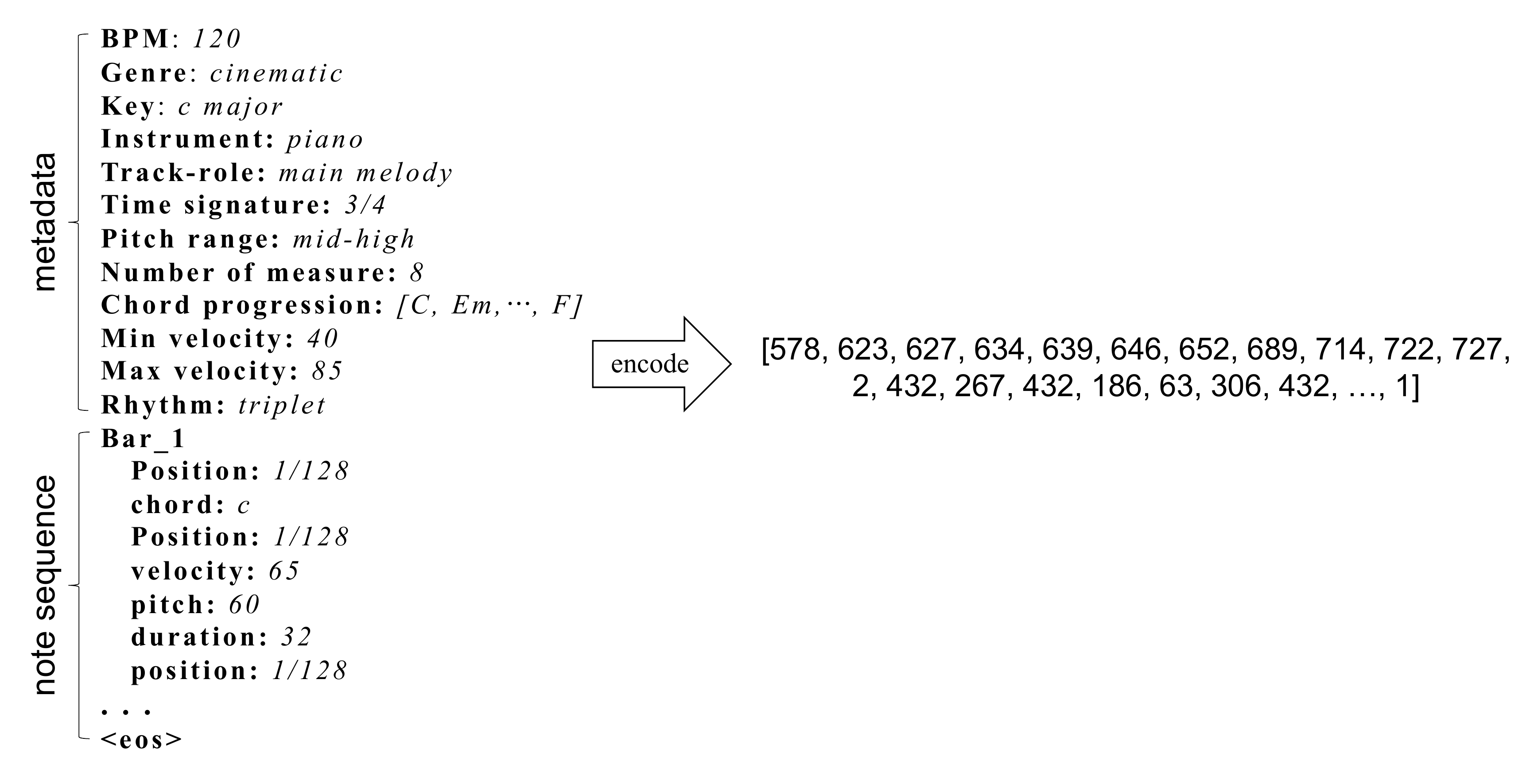}
\caption{An example sequence of the ComMU dataset. We map metadata and note sequence into integer value using encoding dictionary in Table~\ref{table:encoding dict}}
\label{fig:encode}
\end{figure}

Lastly, we upscale representation resolution from 32th note to 128th note. High resolution is effective in representing complex musical expressions such as Arpeggio. See~\ref{appendix:resolution} for experiment about representation resolution. An example of ComMU representation is shown in Figure~\ref{fig:encode}.

\begin{figure}
\centering
\includegraphics[width=1.0\textwidth]{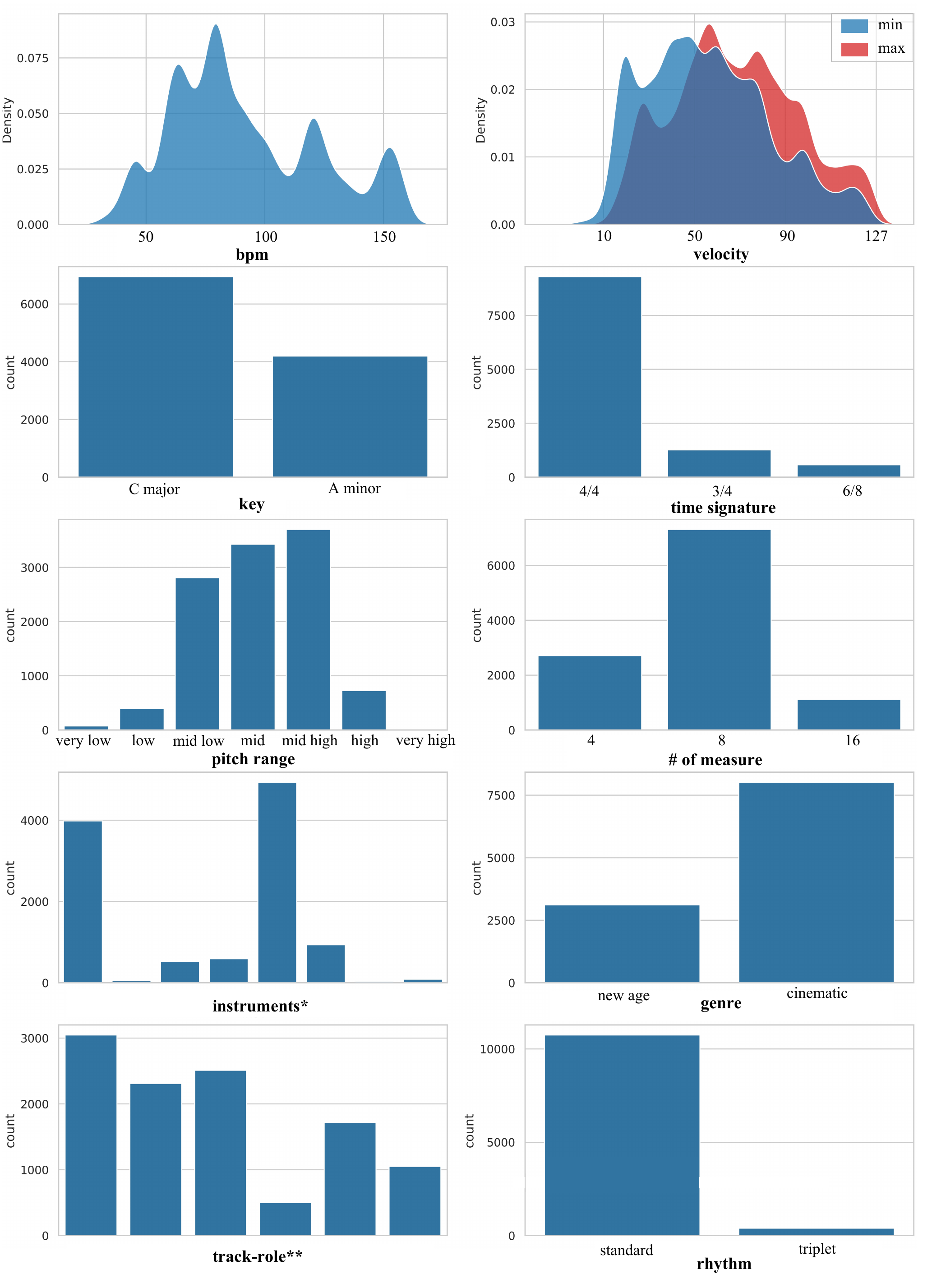}
\caption{Distribution of ComMU metadata. Bpm and velocity are visualized as continuous distributions using kernel density estimation, although they have 40 and 64 discrete values, respectively. \textbf{Instruments*} (in order): keyboard, lead, idiophone, pluck string, string, wind, etc. \textbf{Track-role**} (in order): main melody, sub melody, accompaniment, bass, pad, riff.}
\label{appendix:fig:data_analysis}
\end{figure}
\section{Metadata} \label{appendix:metadata}
    \textbf{BPM.} BPM stands for beat per minute and is a value that represents the tempo of the music. We quantize the BPM into 5 units ranging from 0 to 200 and map them to the nearest integer value.

\textbf{Key.} Key is the group of pitches or scales, that forms the basis of a composition in a classical and the Western pop music. In ComMU dataset, we express the key with 12 root notes (C, C-sharp, D, D-sharp, E, F, F-sharp, G, G-sharp, A, A-sharp, B) and 2 key types (major, minor). Consequently, the vocabulary size of the key token is 24.


\textbf{Instrument.} We encode 128 general MIDI programs(i.e., instruments) into 8 categories — keyboard, lead, idiophone, plucked\_string, string, wind, percussion, and others (ComMU has 37 MIDI programs among 128 programs). The idiophone is an instrument that produces sound primarily through its own vibration, without the use of air flow, strings, membranes, or electricity, such as a bell or a music box. Instruments that are played by plucking the strings are referred to as a plucked string, such as guitar and bass. A technique called plucking involves pulling and releasing the string in order to give it an impulse that makes the string vibrate. A wind instrument is an instrument with a resonator, usually a tube, in which the player blows into (or over) a mouthpiece placed at or near the end of the resonator to cause a column of air to vibrate, such as flute, trumpet, and clarinet. The classification is made depending on how the instruments are played or create the sound as well as how they are used in general in the music industry. For symbolic data representation, this classification is effective in capturing implicit information about music.

\textbf{Time Signature.} Time signature defines how many beats there are in a bar, and which note value the beats are. Each sample is tagged with a time signature of 4/4, 3/4, or 6/8, which are three most frequently used time signatures in pop music.

\textbf{Pitch Range.} We classify samples into 7 different groups of pitch ranges based on the mean pitch of the notes. The sorting criteria is as follows: very low: C-2 - B0, low: C1 - B1, mid low: C2 - B2, mid: C3 - B3, mid high: C4 - B4, high: C5 - B5, very high: C6 - G8.

\textbf{Number of Measures.} The parameter number of measures (i.e., bars) specifies the length of the samples. ComMU typically has 4, 8, or 16 bars of sample, but for incomplete measure, the number of measure can be in decimal (e.g., 8.25).

\textbf{Min Velocity.} This specifies the minimum velocity of the samples.

\textbf{Max Velocity.} This specifies the maximum velocity of the samples.

\textbf{Rhythm.} We divide types of rhythms in ComMU into standard and triplet. Samples with rhythm quantized in straight notes are standard, and other samples with rhythm quantized in triplets or swing notes are triplet.


\section{Additional data analysis} \label{appendix:Additional_Data_Analysis}

Figure~\ref{appendix:fig:data_analysis} shows the distribution of ComMU metadata. BPM, velocity, key, pitch range, and track-role are relatively evenly distributed. Note that genre and key are not severely imbalanced, but the range of choices is small. This is because we first collected the new age and cinematic genres, which are often used in background music. On the other hand, rhythm and time signature are highly imbalanced because 4/4 beat music in standard rhythm is often used in background music. In order to cope with musically complex genres such as jazz in the future, collection of various rhythms and beats is necessary.


Such imbalance in ComMU metadata with few options brings limitations to generating a variety of music. We increased the amount of information in ComMU and decreased some of the imbalances without adding data through the method of key and bpm augmentation~\ref{appendix:A_1}. However, we hope that more diverse data will be added in future studies to increase the capacity of automatic composition and alleviate the stress of data imbalance.

\section{Subjective evaluation} \label{appendix:subjective_evaluation}
    \begin{table} 
  \caption{Standard for subjective metrics.}
  \centering
  \resizebox{0.98\columnwidth}{!}{\begin{tabular}{l|lllll}
  \toprule
    Points & Controllability & Diversity & Humanness & Richness & Overall\\
  \midrule
    10 $\sim$ 9 & in harmony $\sim$ 1 dissonance & Highly varied differences & Very human-like & Very colorful \& rich & Excellent \\
    8 $\sim$ 7 & 2 $\sim$ 3 dissonance & Varied differences & Natural as human & Colorful & Good\\
    6 $\sim$ 5 & 4 $\sim$ 5dissonance & Adequate differences & Average & Average & Average \\
    4 $\sim$ 3 & 6 $\sim$ 7 dissonance & Few differences & Not as natural as humans & Not colorful & Below Average \\
    2 $\sim$ 1 & 8 $\sim$ dissonance & No difference & Not very human-like & Poor & Poor \\
  \bottomrule
  \end{tabular}}
\label{subjective_standard}
\end{table}

In order to find the representation and resolution that best fit our dataset, we conduct a subjective evaluation with internal composers who participated in the production of the ComMU dataset. The criteria standard for subjective evaluation is in Table~\ref{subjective_standard}.
\subsection{Representation: REMI vs. MIDI-like}
REMI representation can capture the sequential information of chord progression. However, even with chord progression prepended, MIDI-like representation is unable to align the chord information with the note sequence. A metrical structure can not be formed without a token implying bar and position. 
This leads to controllability loss and dissonance, resulting in inferior performance. As shown in Table~\ref{t4}, the results generated with the REMI representation are higher in all indices but for subjective diversity.


\begin{table}
    \centering
    \caption{Representation and resolution with the highest score are selected in each of the five metrics. \emph{*score range in 1-10}.}
    \label{t4}
        \begin{tabular}
        {c|ccccc}
        \noalign{\smallskip}\noalign{\smallskip}\hline
        Model & Controllability & Diversity & Humanness & Richness & Overall \\
        \hline
        XL w/ REMI & \textbf{9.49} & 6.48 & \textbf{6.86} & \textbf{6.91} & \textbf{6.90} \\
        XL w/ MIDI-like & 2.53 & \textbf{6.67} & 2.05 & 3.71 & 1.91 \\
        \hline
        XL w/ r32 & 9.51 & 4.14 & 7.23 & 7.11 & 6.99\\
        XL w/ r64 & \textbf{9.67} & \textbf{4.34} & 7.21 & 7.01 & 7.15 \\
        XL w/ r128 & 9.41 & 4.07 & \textbf{7.43} & \textbf{7.37} & \textbf{7.31}\\
        \hline
        \end{tabular}
\end{table}

\begin{figure*}[!t] 
\centering
\subfigure[MIDI with resolution 32]{
\label{fig:resolution32}
\includegraphics[width=0.48\columnwidth, height=20mm]{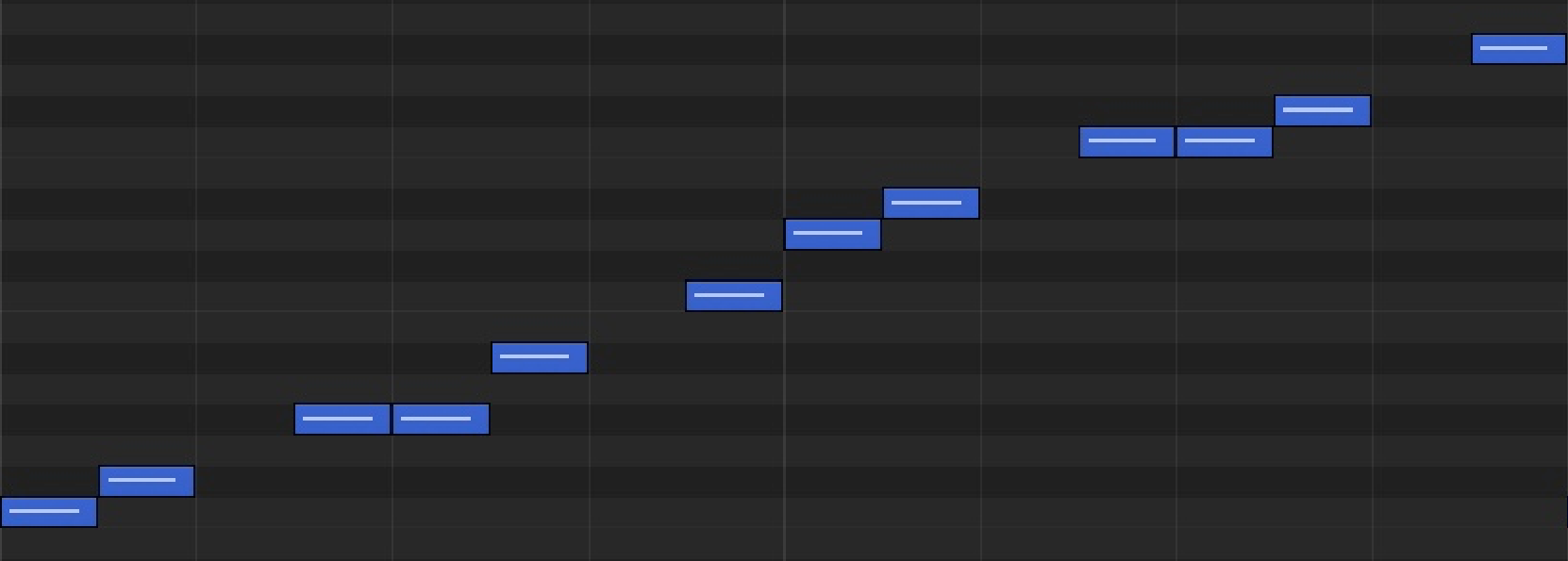}
}
\subfigure[MIDI with resolution 128]{
\label{fig:resolution128}
\includegraphics[width=0.48\columnwidth, height=20mm]{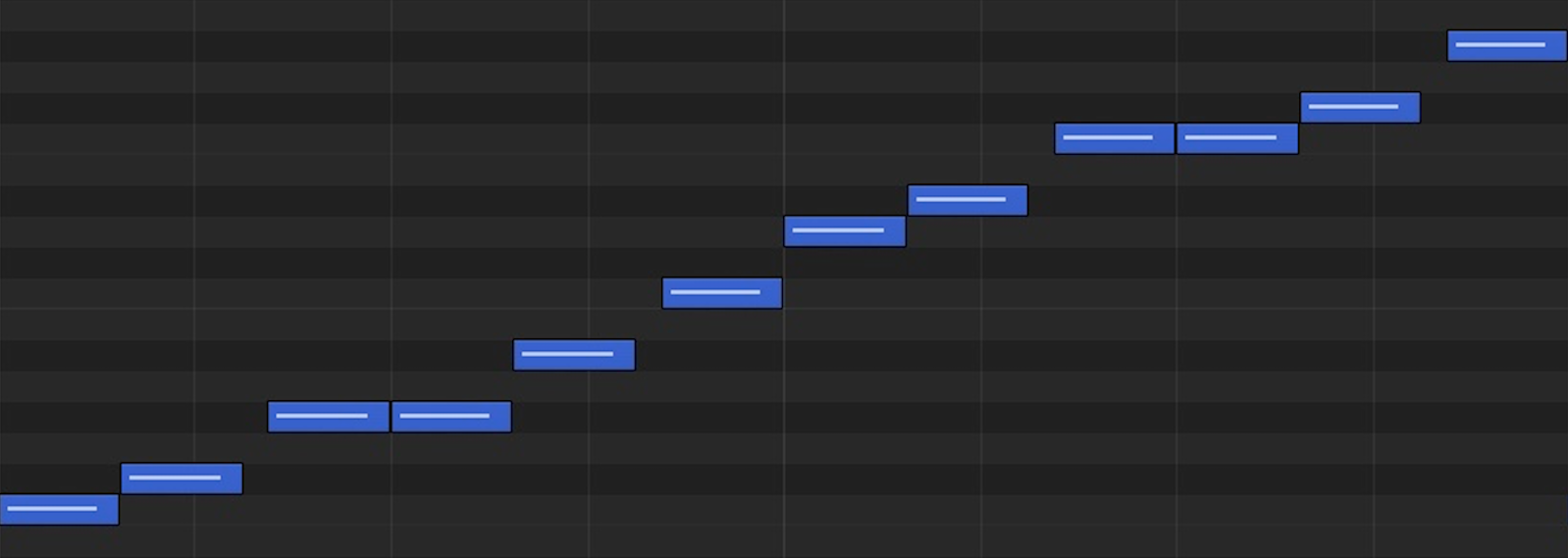}
}
\caption{A sample data of two different resolutions. With low resolution, a note expression may be truncated or placed in an inappropriate position.}
\label{fig:resolution}
\end{figure*}


\subsection{Data representation resolution} \label{appendix:resolution}
We conduct an experiment with representation resolution by manipulating the duration and position of REMI representation to 32, 64, and 128. The resolution is an expression of how many pieces a bar is divided into, and these pieces determine the position and duration of the tokens. The higher the resolution value, the finer the division. 


As shown in Table~\ref{t4}, the result with 128 resolution is slightly lower in terms of controllability.
This is due to an increase in the overall vocabulary size and the occurrence of minute deviations.
However, representation with 128 resolution results in higher scores in overall humanity and richness.
Accordingly, we judge 128 as the reference resolution.
One explanation of the result is that various musical elements such as arpeggio, swing rhythm composed of notes with short duration, and ornaments - trill, appoggiatura, or mordent - can only be expressed through high representation resolution.

In Figure~\ref{fig:resolution}, we compare resolution 32 and 128 in the form of a MIDI. 
The figure illustrates that when the resolution is low, the intervals between notes are far from smooth, resulting in less humanness and richness in sound. 

\section{Additional Experiments} \label{appendix:additional_experiment}
    \begin{table} 
  \caption{We set default values of top-k (K) and temperature ($\tau$) to 32 and 0.95, respectively. Augmentation results come from Transformer-XL. We check diversity (D) and controllability. Controllability metrics include pitch control (CP), velocity control (CV), and harmony control (CH).}
  \vspace{1mm}
  \label{Result}
  \centering
  \begin{tabular}{l|rrr|r}
  \toprule
    &\multicolumn{3}{c|}{Controllability} & Diversity \\
    \multicolumn{1}{c|}{Cases} & \multicolumn{1}{c}{CP $\uparrow$} & \multicolumn{1}{c}{CV $\uparrow$} & \multicolumn{1}{c|}{CH $\uparrow$} & \multicolumn{1}{c}{D $\uparrow$} \\
  \midrule
No aug.  & 0.8433 & 0.8194 & 0.9940 & 0.4386 \\
Key aug. & 0.8270 & 0.8663 & 0.9926 & 0.4238 \\
BPM aug. & 0.8126 & 0.7615 & 0.9914 & 0.4219 \\
Key+BPM aug. & 0.8412 & 0.9102 & 0.9946 & 0.3160 \\
  \midrule
Transformer-XL~\cite{dai2019transformer} & 0.8412 & 0.9102 & 0.9946 & 0.3160 \\
Transformer-GAN~\cite{muhamed2021symbolic} & 0.8502 & 0.8699 & 0.9911 & 0.4097 \\
    \bottomrule
  \end{tabular}
\label{additional_result_table}
\end{table}

\subsection{Augmentation}
We experiment with an ablation study on augmentation to observe its effect on the system. Table~\ref{additional_result_table} shows the increase in controllability of velocity and decrease in diversity of music generated when data augmentation is done. It can be interpreted as that high diversity in the case of no augmentation is due to the lack of data information which results in model under-fitting. We choose the case of key and BPM augmentation as baseline since we are focusing on the controllability of generated music, but we could leverage other cases depending on the purpose. 

\subsection{Different models}
We compare Transformer-XL and Transformer-GAN to confirm the baseline model. Table~\ref{additional_result_table} shows that the controllability of velocity is significantly higher in the Transformer-XL model, but the diversity is higher in Transformer-GAN. This can be interpreted that adversarial loss in Transformer-GAN causes the model to generate more diverse tokens. We use Transformer-XL as a baseline because we focus on the controllability of generated music, but we could leverage other models depending on the purpose.
\section{Training details} \label{appendix:training_detail}
    \begin{wraptable}{r}{0.33\linewidth}
\vspace{-5mm}
\caption{Training details about baseline model.}
 \begin{threeparttable}
\vskip 0.015in
\begin{center}
\begin{small}
\begin{sc}
\begin{tabular}{l|r}
\toprule
\# epochs & 6,000$^{1}$\\
\# GPUs & 4\\
batch size & 256\\
\# layers & 6\\
\# heads & 10\\
dropout rate & 0.1\\
\# parameters & 13,677,310\\
Warmup step & 100\\
Learning rate & 0.004\\
scheduler & inv\_sqrt\\ 
optimizer & adam\\
\bottomrule
\end{tabular}
\end{sc}
\begin{tablenotes}
  \item[1] We got the lowest validation nll loss at 6,000 epochs.
  \end{tablenotes}
\end{small}
\end{center}
  \end{threeparttable}
\vskip -0.7in
\label{table:training_detail}
\end{wraptable}

In this section, we describe critical hyperparameters and resources used here. Training details are descirbed in Table~\ref{table:training_detail}.  The CPU used here is Intel core X-series i9-10900X, whose base clock speed is 3.70 GHz, and its cache is 19.25 MB Intel smart cache. Further, we utilized 4 GPUs (GeForce RTX 3090 D6X).

\section{Societal impact} \label{appendix:societal_impact}


ComMU can not only complete combinatorial music generation task but has also freed up the possibility of building a model that back-tracks the metadata from a given piece of music, and even study the correlation between each metadata.

However, it is crucial to recognize the social and ethical challenges of automatic music generation. When used outside the boundaries of commercial applications specified in the license, infringement of copyright and licensing may become an issue. Namely, illegal redistribution of music and dataset via sound sample platforms such as Splice may occur. It is also notable that music generated through our dataset, while large in sample size, cannot include the entire music culture, possibly dominated by the Western genres and instruments. As the scope of the music which can be generated is strictly limited to the dataset itself, it is impossible to not marginalize any musical culture.
\section{Obtained consent from composers} \label{appendix:obtained_consent}

As aforementioned under subsection~\ref{subsection:dataset_collection}, MIDI samples with metadata for constructing ComMU dataset were manually produced by 14 professional composers. We obtained consent from the composers when we hired them for their work. Thus, the ComMU dataset is \emph{work made for hire} and it belongs to the company.

\section{Amazon mechanical turk} \label{appendix:amazon_mturk}
\begin{figure}
\centering
\includegraphics[width=1.0\textwidth]{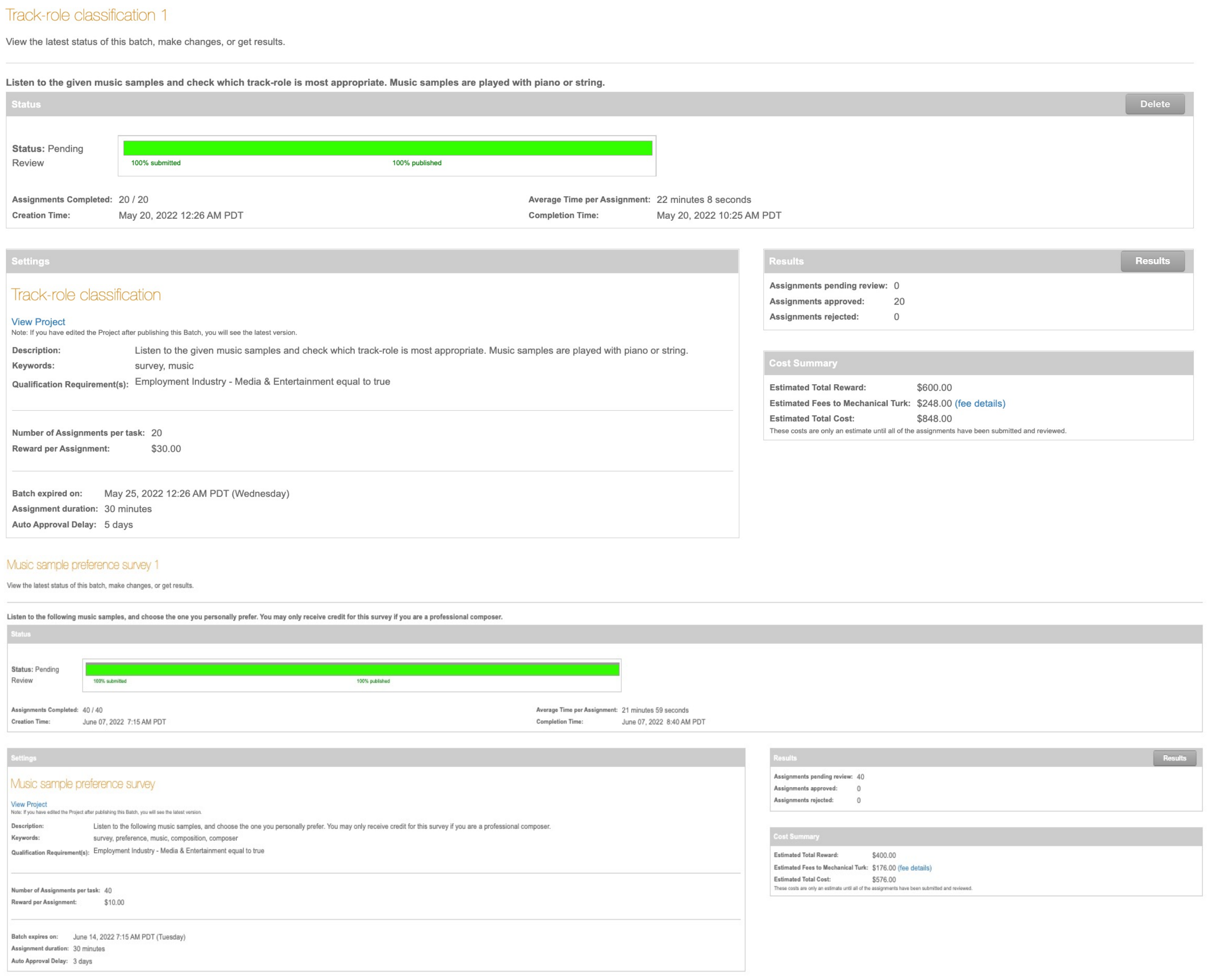}
\caption{Human intelligence tasks(Hits) template for track-role classification and music samples preference survey.}
\label{appendix:fig:mturk}
\end{figure}

\begin{figure}
\centering
\includegraphics[width=1.0\textwidth]{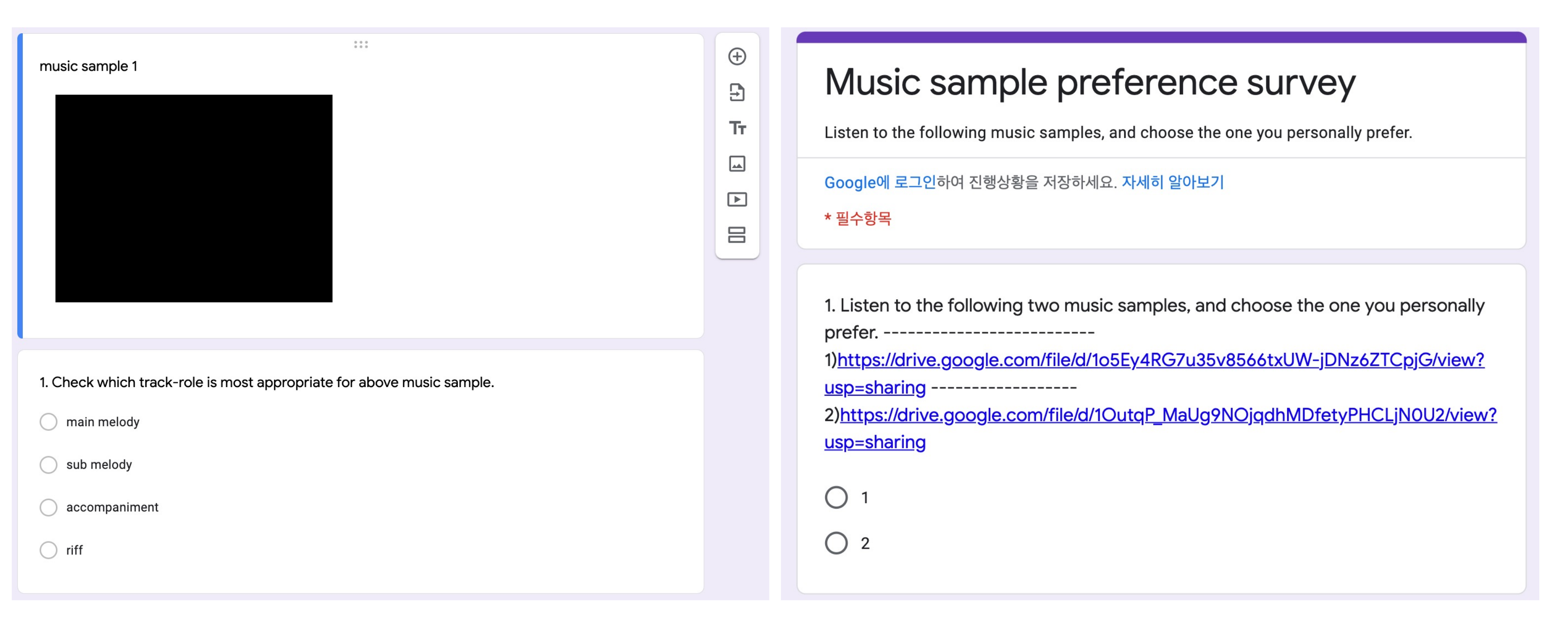}
\caption{Example of google form for track-role classification and music samples preference survey.}
\label{appendix:fig:form}
\end{figure}

    We use amazon mechanical turk for human studies: preference survey for measuring fidelity(Section~\ref{results}), and track-role classification for validating the advantage of applying unique metadata(Section~\ref{5.2_track-role}). Figure~\ref{appendix:fig:mturk} specifies the details about the surveys such as estimated time required to complete the survey and reward for their participation. Figure~\ref{appendix:fig:form} shows specific instructions delivered to subjects in google form. 
\section{Music terminology} \label{appendix:music_term}
    \textbf{Pitch.} A Pitch is a perceptual characteristic of sounds that enables their classification on a scale of frequency, or more commonly~\cite{klapuri2006introduction}, pitch is the ability to identify sounds as "higher" and "lower" in the sense associated with musical melodies~\cite{plack2006pitch}.

\textbf{Note.} A note is a symbol for a musical sound in music. In musical notation, notes can stand in for the pitch and duration of a sound and notes are fundamental units of music.

\textbf{Key.} In music, the key is a group of pitches or scales, that forms the basis of a musical composition in a classical and the Western pop music.

\textbf{Harmony.} A Harmony is a process of playing various notes in a way that makes them sound right when played together. Individual sounds are combined or composed into larger units or compositions through the process of harmony~\cite{lomas2022harmony}.

\textbf{Chord.} A chord is any harmonic grouping of pitches or frequencies made up of several notes that are perceived as sounding simultaneously~\cite{benward2003music}.

\textbf{Root.} A root is the base note of a key, scale or chord. For example, in the key of C major, the root note is C.

\textbf{MIDI.} MIDI (i.e., Musical Instrument Digital Interface) is a technical standard that describes a communications protocol, digital interface, and electrical connectors that connect a wide variety of electronic musical instruments, computers, and related audio devices for playing, editing, and recording music\cite{rothstein1992midi}.

\end{document}